\documentclass[aps,prc,twocolumn,nofootinbib,superscriptaddress]{revtex4-1}

\usepackage{bm}        % for math
\usepackage{xcolor}

\usepackage{dcolumn}   % needed for some tables
\usepackage{multirow}

\usepackage{amsmath}
\usepackage{amsfonts}
\usepackage{amssymb}
\usepackage{makeidx}
\usepackage{graphicx}
\usepackage{anysize}
\usepackage{slashed}
 \usepackage{bbm}
 \usepackage{physics}
 \usepackage{comment}

\usepackage{xcolor}

\usepackage{epstopdf}

\usepackage{soul}
\usepackage[normalem]{ulem} %for \sout{} and \uline{}
\usepackage{cancel}

\usepackage{colortbl} %colors to hlines

\usepackage{booktabs}
\usepackage{hyperref}

\usepackage{tikz}
\usetikzlibrary{svg.path}
\definecolor{orcidcol}{HTML}{A6CE39}
\tikzset{
  orcidlogo/.pic={
    \node[anchor=south west,inner sep=0] (img) at (0,0) {};
    \fill[orcidcol] (0.12,0.12) circle (0.12);
    \node[white] at (0.12,0.12) {\fontsize{1.5}{2}\selectfont\bfseries iD};
  }
}
\newcommand{\orcid}[1]{\href{https://orcid.org/#1}{\mbox{\hspace{0.cm}\tikz[baseline=-0.1ex]{\pic{orcidlogo};}}}}

\newenvironment{mymathbox}
{\par\smallskip\centering\begin{lrbox}{0}%
\begin{minipage}[c]{0.8\textwidth}}
{\end{minipage}\end{lrbox}%
\framebox[0.9\textwidth]{\usebox{0}}%
\par\medskip
\ignorespacesafterend}
\newcommand{\bb}{\begin{mymathbox}}
\newcommand{\eb}{\end{mymathbox}}

\newcommand{\be}{\begin{equation}}
\newcommand{\ee}{\end{equation}}
\newcommand{\ba}{\begin{eqnarray}}
\newcommand{\ea}{\end{eqnarray}}

\newcommand{\nk}{{\bf      k}}
\newcommand{\np}{{\bf      p}}
\newcommand{\nq}{{\bf      q}}

\newcommand{\npsi}{{\bf \npsi}}

\newcommand{\de}{\text{d}}

\newcommand{\non}{\nonumber}

\newcommand{\bma}{\begin{pmatrix}}
\newcommand{\ema}{\end{pmatrix}}

\begin{document}

\title{Interplay between final-state nucleon distortion, nuclear transparency and intranuclear cascade models in neutrino event generators}

\author{R.~Gonz\'alez-Jim\'enez
\orcid{0000-0002-0492-0619}}
\email{raugj@us.es}
\affiliation{Departamento de Física Atómica, Molecular y Nuclear, Universidad de Sevilla, 41080 Sevilla, Spain \looseness=-2}

\author{A.~Nikolakopoulos
\orcid{0000-0001-6963-8115}}
\email{anikolak@uw.edu}
\affiliation{University of Washington, Seattle, WA 98195, USA \looseness=-2}

\author{J.~McKean
\orcid{0009-0005-6100-6195}}
\email{mckean.jake.42u@st.kyoto-u.ac.jp}
\affiliation{Kyoto University, Department of Physics, Kyoto, Japan \looseness=-2}

\author{K.~Niewczas
\orcid{0000-0001-9109-197X}}
\affiliation{\'Ecole polytechnique, IN2P3-CNRS, Laboratoire Leprince-Ringuet, F-91120 Palaiseau, France \looseness=-2}
\affiliation{Department of Physics and Astronomy, Ghent University, B-9000 Gent, Belgium \looseness=-2}

\author{T.~Franco-Munoz
\orcid{0009-0000-3871-4752}}
\affiliation{Department of Physics and Astronomy, Ghent University, B-9000 Gent, Belgium \looseness=-2}
 
\author{J.~Garc\'ia-Marcos
\orcid{0009-0003-2753-1864}}
\affiliation{Department of Physics and Astronomy, Ghent University, B-9000 Gent, Belgium \looseness=-2}
\affiliation{Grupo de F\'isica Nuclear, 
Departamento de Estructura de la Materia, F\'isica T\'ermica y Electr\'onica, 
Facultad de Ciencias F\'isicas, Universidad Complutense de Madrid and IPARCOS, CEI Moncloa, Madrid 28040, Spain \looseness=-3}

\author{M.~Hooft
\orcid{0000-0002-6368-4820}}
\affiliation{Department of Physics and Astronomy, Ghent University, B-9000 Gent, Belgium \looseness=-2}

\author{N.~Jachowicz
\orcid{0000-0003-1168-0745}}
\affiliation{Department of Physics and Astronomy, Ghent University, B-9000 Gent, Belgium \looseness=-2}

\author{G.~D.~Megias
\orcid{0000-0002-0890-1343}}
\affiliation{Departamento de Física Atómica, Molecular y Nuclear, Universidad de Sevilla, 41080 Sevilla, Spain \looseness=-2}

\author{J.~M.~Ud\'ias
\orcid{0000-0003-3714-764X}}
\affiliation{Grupo de F\'isica Nuclear, 
Departamento de Estructura de la Materia, F\'isica T\'ermica y Electr\'onica,
Facultad de Ciencias F\'isicas, Universidad Complutense de Madrid and IPARCOS, CEI Moncloa, Madrid 28040, Spain \looseness=-3}

\date{\today}

\begin{abstract}
We present an overview of the relativistic distorted-wave impulse approximation (RDWIA) for lepton–nucleus scattering, with emphasis on its use as an elementary-vertex model in event generators. In the intermediate-energy regime, neutrino–nucleus event generators typically follow a two-step scheme: an elementary neutrino–hadron interaction is first generated, and the outgoing hadrons are then propagated through the nuclear medium using an intranuclear cascade (INC) model. Using exclusive  and inclusive $(e,e'p)$ and $(e,e')$
data for several nuclei, we assess how nuclear effects should be distributed between these two stages. In particular, we identify which effects should be included in the elementary-vertex description and which can be treated more effectively by the INC, while addressing the consistency required between both components.

We also discuss nuclear transparency in quasielastic scattering. We show that it is directly connected to $(e,e'p)$ data below the two-nucleon knockout threshold; and depends on the full nuclear density distribution and the initial position of the nucleon, therefore, it varies substantially with the shell from which the nucleon is removed. 
We demonstrate that this shell dependence can be incorporated straightforwardly into the INC component of event generators.
\end{abstract}

\maketitle

\section{Introduction}

Event generators are a necessary tool in long baseline neutrino-oscillation experiments (T2K~\cite{T2Kweb}, HK~\cite{HKweb}, SBN~\cite{SBNweb}, MicroBooNE~\cite{MicroBooNEweb}, NOvA~\cite{NOvAweb} and DUNE~\cite{DUNEweb}) to connect theoretical model predictions with the different event topologies in the detectors. In particular, they are used to reconstruct the energy of the neutrino, which is a key ingredient for oscillation analyses~\cite{Alvarez-Ruso18, Dolan26b}.

Most of the current event generators assume a two-step approximation: (1) an elementary-vertex interaction, and (2) a subsequent propagation through the nucleus of the hadron(s) produced in the previous step.
In the intermediate energy regime typical of the above mentioned oscillation experiments, i.e. from a few hundreds MeV to a few GeV, the main elementary-vertex interactions, also called reaction mechanisms, are quasielastic (QE) scattering, two-body mechanisms leading to two particle-two hole (2p-2h) final states, single-pion production (SPP), multipion production, production of other mesons, and deep-inelastic scattering (DIS).
There are many different nuclear frameworks to describe the various reaction mechanisms, such as infinite nuclear matter approaches, shell models or {\it ab initio} approaches, a comprehensive review can be found in \cite{Alvarez-Ruso25}. 
On the other hand, hadron propagation is usually handled using classical intranuclear cascade (INC) models, as is the case for the Monte Carlo neutrino event generators NEUT~\cite{Hayato21}, NuWro~\cite{NuWro-web, Niewczas19}, GENIE~\cite{Andreopoulos10} and ACHILLES~\cite{Isaacson23}; other options exist, like GiBUU~\cite{Buss12}, based on kinetic transport theory.

This factorization into elementary vertex and subsequent evolution of the hadron system is common practice in DIS, where a {\it hard} lepton-quark interaction that completely defines the inclusive cross section~\footnote{By inclusive cross section we refer to the case in which only the scattered lepton is detected, so the cross section only depends on leptonic variables and would get contributions from all open channels.} is followed by a hadronization process. 
In this work, we focus on an energy region below DIS, such that the relevant nuclear degrees of freedom are nucleons, not quarks. We will use QE scattering as a driver of our discussion, but the same ideas would easily be applicable to other reaction mechanisms, like SPP or 2p-2h induced by meson-exchange currents (MEC).

QE scattering is incorporated in the two-step framework as a primary interaction with a single nucleon, which is promoted to a continuum state with (asymptotic) momentum $\vb{p}_N$; this is determined by the elementary-vertex model with differential cross section $\frac{\de\sigma}{\de\nk_f \de\vb{p}_N}$. Then, one includes the probability for rescattering, so that a different and possibly more complex final-state is produced, which we denote as a shorthand simply by $\vb{p}_F$, even though this can be a multiparticle state. 
Hence, the cross section for production of the specific hadronic final-state is
\begin{equation}
    \frac{\de\sigma}{\de\nk_f \de\vb{p}_F} = \int \de\vb{p}_N \frac{\de\sigma}{\de\nk_f \de\vb{p}_N} 
    F(\vb{p}_F \vert \vb{p}_N ).\label{eq:twostep}
\end{equation}
The last factor is the conditional probability density function for production of a final state $\vb{p}_F$ given $\vb{p}_N$. This probability for rescattering is computed classically with the INC model, hence, one has the unitarity condition
\ba
  \int \de\np_F\, F(\np_F|\np_N) = 1\,. \label{eq:unitarity}
\ea

Then, the inclusive cross section is determined solely by the cross section used as input, since the integral over all final-states is
\ba
 \frac{\de \sigma}{\de \nk_f} &=& \int{\de \np_F \frac{\de\sigma}{\de\nk_f \de \np_F}}\non\\ 
    &=& \int{ \de\vb{p}_N \frac{\de\sigma}{\de\nk_f \de\vb{p}_N} \int{\de \np_F F(\np_F | \np_N)}  },\label{eq:inclusive}
\ea
and the rightmost integral is one by Eq.~\eqref{eq:unitarity}.
In other words, the INC distributes the flux from the elementary vertex in a unitary way into a multiplicity of final states, but, if one integrates over all hadron variables, the inclusive cross section provided by the elementary-vertex model is recovered.

\begin{figure}[htb]
\centering
\includegraphics[width=.5\textwidth,angle=0]{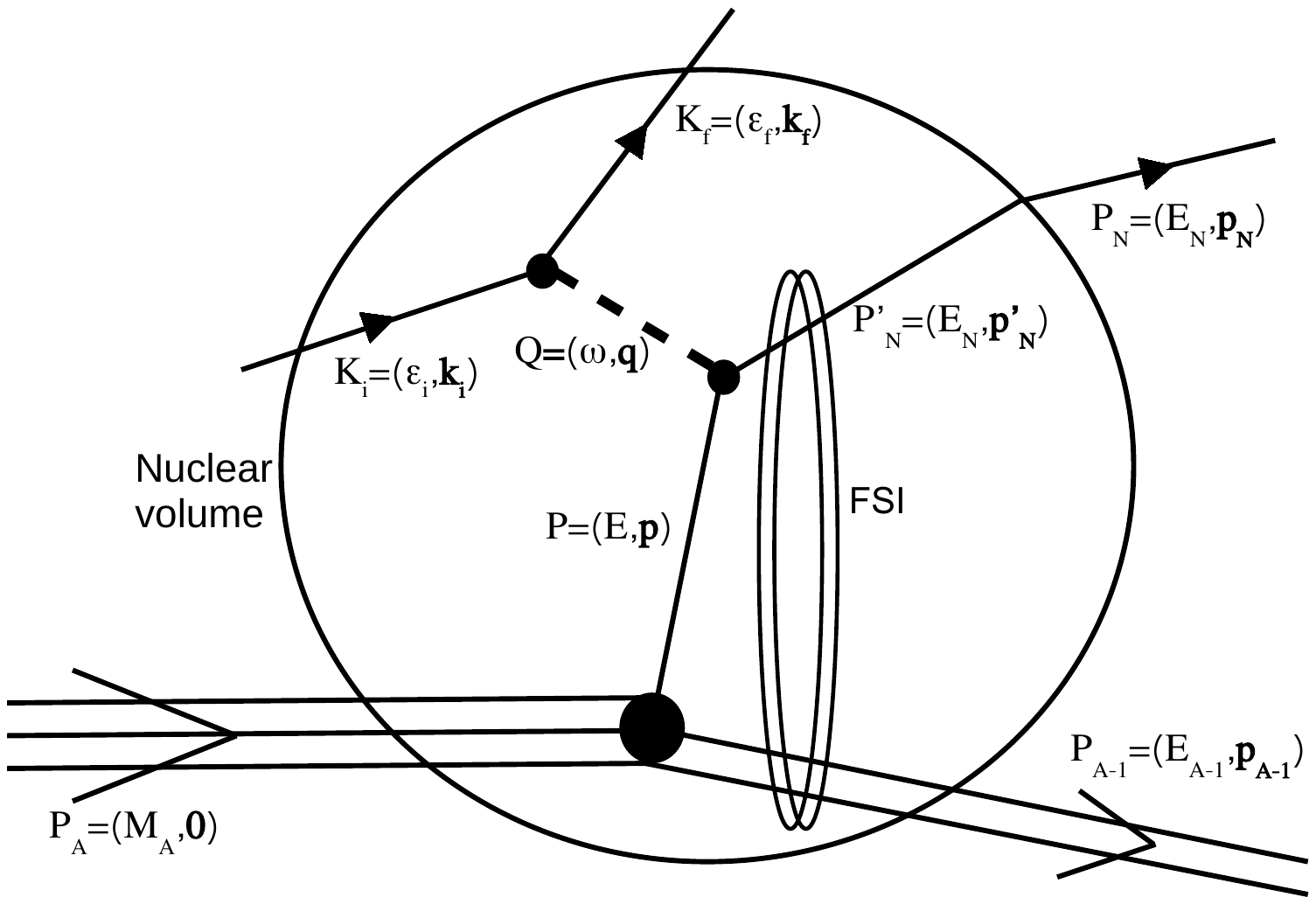}
\vspace{-1cm}
\caption{Quasifree one-nucleon knockout induced by lepton scattering in the one-boson exchange and impulse approximations.}
\label{fig:QE-scattering}
\end{figure}

Thus, in this two-step description, the INC step cannot change the inclusive cross section at all, meaning that this approach can only reproduce the inclusive results  if we use an elementary vertex that by itself predicts adequately the measured inclusive cross section. 
While this approach is common, it is not a priori obvious that it should be possible to describe the inclusive cross section by considering only the simple  interaction mechanisms, rather than requiring a full description for all exclusive channels. 
However, this seems to be the case.
In the region of the quasielastic peak, especially for the longitudinal electromagnetic response, the inclusive cross section can be described by considering only the quasifree one-nucleon knockout process~\cite{Franco-Munoz23,Cavallin26,Jachowicz21,Lovato25,Niewczas26}
\ba
\ell + {}^A\text{X}\longrightarrow \ell' + N + {}^{A-1}\text{Y}\,.
\ea
In the computation of this process, the energy transferred to the target is mostly transferred into kinetic energy of a single struck nucleon (apart from binding energy and recoil energy of the ${}^{A-1}$Y system). The mechanism is sketched in Fig.~\ref{fig:QE-scattering}. 
While this approach is successful in describing the inclusive cross section around the QE peak, robust estimates of the actual contribution of quasifree one-nucleon knockout near the QE peak are around $20\%-80\%$ of the total inclusive rate~\cite{Kelly96}. 
This fraction is closely related to the concept of `nuclear transparency', which is discussed in this work.
Its value depends on the particular kinematics and density of the nucleus. 
The key point is that it is far from 100\% and $A$-dependent, which motivates a comprehensive study of nuclear transparency across different nuclear models and INC implementations.

The two-step approach is further supported by the observation that, above the Pauli blocking region and for the longitudinal sector, scaling works remarkably well in the QE peak~\cite{Day90,Donnelly99a,Donnelly99b,Maieron02}. 
The occurrence of this phenomenon provides strong evidence that electron-nucleus scattering in the QE peak is primarily driven by the elastic interaction of the electron with a single nucleon.
The scaling property is derived under the quasifree assumption that a single nucleon absorbs the transferred energy and momentum. 
Conversely, if it was necessary to track how energy is distributed over the full hadronic final state, scaling would break down (in an $A$-dependent manner). This is for example the case for the transverse response, where two-body interaction mechanisms break scaling~\cite{Donnelly99b}.
The case of final-state interactions included through the two-step approach is different, the rescatterings responsible for the reconfiguration of the hadronic final state  are detached, in space and time, from the initial-vertex interaction.

The following questions come up naturally when assuming this two-step approximation:
\begin{itemize}
    \item Which nuclear effects should be included in the initial interaction model and which ones in the hadron propagation?
    \item What level of consistency between the elementary-vertex and INC models is required?
    \item Which experimental data can be used to benchmark and refine these models?
\end{itemize}

In this work, we assess these questions by comparison to theoretical calculations and electron scattering data. 
We make a distinction between elastic and inelastic final-state interactions. The former are naturally included in the elementary vertex model if the bound- and continuum-states are treated consistently.
We then consider different ways to define the nuclear transparency and their utility in different scenarios.
We stress that transparency is a model-dependent quantity. Therefore, when using it to constrain an INC, one should ensure consistency with the model used as input to the INC.
RDWIA calculations show that nuclear transparency depends strongly on the orbital from which the nucleon is knocked out. We show that this effect has a clear classical interpretation, it can be reproduced by INC models and straightforwardly implemented in event generators.
Finally, a comparison with exclusive $(e,e' p)$ data reveals where and to what extent the two-step approach breaks down, which sets a benchmark for the further improvement of INC models.

Although the primary objective of this work is to guide improvements in Monte Carlo neutrino event generators, we do not present neutrino-nucleus cross-section results here. Instead, we focus on higher-precision electron–nucleus scattering data to constrain nuclear effects, which are largely probe-independent for a given target nucleus.

This paper is organized as follows. In Sec.~\ref{sec:elementary-vertex}, we describe a model for the elementary vertex; namely, the relativistic distorted-wave impulse-approximation (RDWIA). In Sec.~\ref{sec:Transparency}, we discuss the concept of nuclear transparency within the RDWIA and INC models. In Sec.~\ref{sec:Results} we present results for inclusive and exclusive electron scattering data, and connect these with the transparency predictions from the INC and the RDWIA. Finally, we summarize our conclusions in Sec.~\ref{sec:conclusions}.

\section{Elementary-vertex models}\label{sec:elementary-vertex}

We describe two different scenarios within the framework of the RDWIA approach: exclusive one-nucleon knock-out and inclusive cross sections. 
Then, we discuss other features of this approach.

RDWIA is based on the following basic ingredients: 

i) Impulse approximation, i.e., the virtual boson couples to only one nucleon that is knocked out, and may be detected.  

ii) The initial bound state is described by relativistic Dirac spinors. In our case, we use an independent-particle shell model (IPSM), specifically the relativistic mean-field (RMF) model of Refs.~\cite{Horowitz81,NLSH}. Then, we account for effects beyond the IPSM, mainly, by using phenomenological spectroscopic factors (SFs).

The model used here is exactly the same as the one described in Refs.~\cite{Franco-Munoz23,Franco-Munoz25} (with the one-body current only). 
Hence, all calculations presented here introduce spectroscopic factors in the overlap of the A-1 and A systems.
The spectroscopic factor is the number of nucleons with a given orbital and angular momentum that can be knocked out for a given missing energy [defined in Eq.~\eqref{eq:Emiss}]. This corresponds to the normalization of the overlap or quasiparticle function, with the corresponding quantum numbers and in the energy range considered. 
Within an independent-particle model, these spectroscopic factors correspond to  the full occupancy of the shell. 
Effects beyond mean field drive part of the strength of these spectral amplitudes towards higher missing energies and missing momentum, and provide also some energy width to what would otherwise be Dirac delta functions, for the states appearing at missing energies below the two-nucleon knockout threshold. 

And iii), the hadronic final state is the ${}^{A-1}Y$ nucleus plus the scattered nucleon, which is described as the solution in the continuum of the wave equation with relativistic potentials.

In the RDWIA, the hadron current computed in momentum space reads:
\begin{flalign}
   J^\mu(\nq,\np_{as};S,\alpha) =
   \int d\np\, \overline\Psi(S;\np',\np_{as}){\cal O}^\mu\Psi(\alpha;\np)\,,\label{hadron-current}
\end{flalign}
where ${\cal O}^\mu$ is the QE operator, $\Psi(\alpha;\np)$ is a momentum-space spinor representation of the hole state (overlap of the A-1 and A systems),
and $\alpha$ encodes all quantum numbers for the hole state.
The scattering state is given by the spinor $\Psi(S;\np',\np_{as})$, with well defined asymptotic momentum $\np_{as}$ and spin projection $S$; again, we chose to represent it in p-space and evaluate it at $\np'=\nq+\np$.
To compute the matrix element $J^\mu$, we perform a weighted average over all contributing $\np$ (and hence $\np'$), this is the meaning of the integral over $\np$ in Eq.~\eqref{hadron-current}. 

\subsection{Exclusive one-nucleon knockout cross section}\label{subsec:exclusive}

In exclusive ${}^A$X$(e,e'N){}^{A-1}$Y experiments, the process results in a lepton detected in coincidence with a knockout nucleon and energy-momentum conservation provides certainty that the detected nucleon is the only knocked-out nucleon.  
This is ensured by restricting the missing energy $E_m$ to be in narrow windows that select specific states and by keeping $E_m$ below the two-nucleon knockout threshold. The missing energy is defined as
\ba \label{eq:Emiss}
E_m = M_Y + M - M_X = \omega - T_p - T_Y,
\ea
with $M$, $M_X$, $M_Y$ the masses of nucleon, target and residual system; and $T_{p,Y}$ the kinetic energies of the detected proton and residual system. 

To describe these exclusive processes, the nucleon in the continuum can be described using relativistic complex-optical potentials (ROP); in our case, as in most recent works, taken from Refs.~\cite{Cooper93,Cooper09}. These ROP were extracted by fitting elastic proton-nucleus scattering data $^A$X$(p,p')^A$X.
The final state in the exclusive ${}^A$X$(e,e'N){}^{A-1}$Y experiment for the knocked out nucleon is analogous to elastic nucleon-nucleus scattering, 
in both cases, the kinematics guarantees that the nucleons have scattered elastically off the nucleus.
It is thus a sensible choice to use that same optical potential to describe the proton in the final state of $(e,e'p)$ processes, that in this way is, to a large extent, constrained by experiment.

A complex optical potential does not preserve the flux of particles coming out the nucleus, the imaginary term introduces a reduction of the outgoing flux of particles, that is, of knocked-out nucleons.
In this way, the optical potential takes into account that a fraction of knocked-out nucleons experience inelastic interactions 
(knocking out additional nucleons, but also exciting the residual system in any other conceivable way).
The fact that the outgoing nucleon is sometimes lost into inelastic channels is captured by the imaginary part of the optical potential.
If the nucleon scatters inelastically with the target, it will not contribute to the elastic $^A$X$(p,p')^A$X measurement, which was used to extract the optical potential. 
On the other hand, the real part of the potential accounts for the elastic interactions that the nucleon undergoes in the scattering process. 
In the context of event generators, these `elastic FSI' are also referred to as `FSI at the vertex' or, in more general contexts, as `nucleon distortion'.

Roughly, the elastic interactions (real potential) change the shape and position of the cross section, while the imaginary potential reduces its overall normalization.
This relatively simple picture of the role of real and imaginary components of the potential gets more involved if one inspects their actual effects in the one-nucleon knockout cross section.

The inelasticity in a partial wave channel is determined by the imaginary part of the phase shift. When solving the Dirac equation, variations in the shape of the real (or imaginary) potential affect both the real and the imaginary parts of the phase shift. For nucleon knockout, the situation is even more complicated, since in this case the full wave function, and not only the phase shifts, enters the overlap integrals of Eq.~\eqref{hadron-current}.
The consequence is that both the real and imaginary parts of the potentials affect the shape and magnitude of the cross section, it will be shown in the results section.

Comparisons with exclusive $^A\text{X}(e,e'p)^{A-1}\text{Y}$ data have shown that the RDWIA works fairly well in a wide energy range and for several nuclei~\cite{Udias93, Meucci01,Kelly05,Butkevich25,Choi21}. 
In what follows, we will refer to the approach based on the use of complex relativistic optical potentials as the `ROP model'.

\subsection{Inclusive cross section}\label{subsec:inclusive}

For inclusive cross sections, the final state is not fixed (or restricted to the elastic propagation of the knocked-out nucleon) as it was for the exclusive case previously described. 
Thus, the one-nucleon knockout contribution to inclusive scattering (that is, not counting the low energy transfer cross section where the residual system is not broken) is described within the same framework used for the quasifree one-nucleon knockout scattering process, 
but with a real potential~\cite{Gonzalez-Jimenez19, Franco-Munoz23}, 
often, the real part of the complex optical potential is used~\cite{Butkevich07,Kim03}. 
Another common approach is to use the same energy independent potential used to describe the bound state~\cite{Pandey15,Caballero05}; another one is the relativistic Green function approach~\cite{Meucci03}, where one uses the same complex optical potential for the exclusive and inclusive cases, but in the inclusive case a re-summation of all open channels is performed so the flux lost to inelastic channels is recovered.

In this work, we will use the real part of the complex relativistic optical potentials from~\cite{Cooper09} and the energy-dependent relativistic potential derived from the self-consistent solution of the RMF problem of~\cite{Gonzalez-Jimenez19}, we will refer to them as `r(ROP)' and `EDRMF', respectively.
Hence, the inclusive cross section is computed by integrating the $(e,e'N)$ result with a real potential over the final-nucleon variables.

\subsection{Plane waves, orthogonality and FSI at the elementary vertex}\label{subsec:RPWIA}

In the context of $(e,e'p)$ and $(e,e')$ studies,  
final-state interactions (FSI) are often considered to be included when the outgoing nucleon is described by a distorted wave obtained by solving a wave equation with nuclear potentials, as in the RDWIA. Conversely, when the final nucleon is described by a plane wave, i.e., a solution of the free wave equation, as in the relativistic plane-wave impulse approximation (RPWIA), FSI are considered to be absent.

If hadronic initial ($I$) and final ($F$) states are not orthogonal, then spurious contributions appear in the cross section because $\langle I | F \rangle \neq 0$, which means that even in the absence of an interaction the initial state can end up in the final state. 
This is a serious issue e.g. within the RPWIA, as shown in~\cite{Gonzalez-Jimenez19,Nikolakopoulos19}. 
Orthogonality, as gauge invariance~\cite{Nikolakopoulos25}, is ensured if the states are eigenstates of the same Hamiltonian, in other words, solutions of the same wave equation.

In summary, using a plane wave for the scattered nucleon and a shell-model bound state for the initial one not only implies that elastic FSI is not included, it has other negative effects. 
Only when the overlap between initial and final states is close to zero, which happens when the momentum of the knocked-out nucleon is much larger than the typical Fermi momentum of bound nucleons, the spurious contributions due to non orthogonality of states~\cite{Gonzalez-Jimenez19} are close to zero. And only when the nucleon wavelength ($\lambda$) is much shorter than the average distance between nucleons in the nucleus, one can ignore the elastic FSI~\footnote{Considering an average distance between nucleons in the nucleus of $1.8$ fm, one expects elastic FSI to be negligible for nucleon momentum $k\gg700$ MeV.}.

\section{Nuclear Transparency}\label{sec:Transparency}

In general terms, nuclear transparency refers to the fraction of nucleons that can travel throughout the nucleus without interacting with the nuclear system. In QE electron-nucleus scattering, this refers to the fraction of nucleons that, after being knocked out by the exchanged photon, travels the nucleus without interacting inelastically with the nuclear system.
While this classical concept is easy to understand, it is not straightforward to find a measurable definition to quantify it.
Hence, there may be different predictions of transparency, linked to different ways to define it.

In electron scattering experiments, nuclear transparency has generally been defined as the ratio of $(e,e'p)$ cross section data to the cross section computed with a plane-wave impulse-approximation model (PWIA), i.e., $\sigma_{\text{data}}/\sigma_\text{PWIA}$, see~\cite{Niewczas19} for a recent review.

In a recent work~\cite{Dytman26}, a more data-driven definition of transparency has been proposed; in this case, both numerator, $(e,e'p)$, and denominator, $(e,e')$, are defined using experimental yields; though, the use of model-dependent simulations was unavoidable for the construction of the denominator of the ratio. 
This way of analyzing and presenting experimental data makes it difficult to make clean theory-data comparisons. 
We believe that the most valuable output of an experiment is a model-independent cross section, as it permits the unambiguous construction of diverse observables, including different measures of `transparency'.

In the following, we discuss the nuclear transparency in the INC; then, we show results in the RDWIA, which allows us to demonstrate how the classical concept of transparency breaks down.

\subsection{Transparency in INC models}

\begin{figure*}
\centering
\vspace{-0.5cm}
\hspace{-1cm}
\includegraphics[width=.55\textwidth,angle=0]{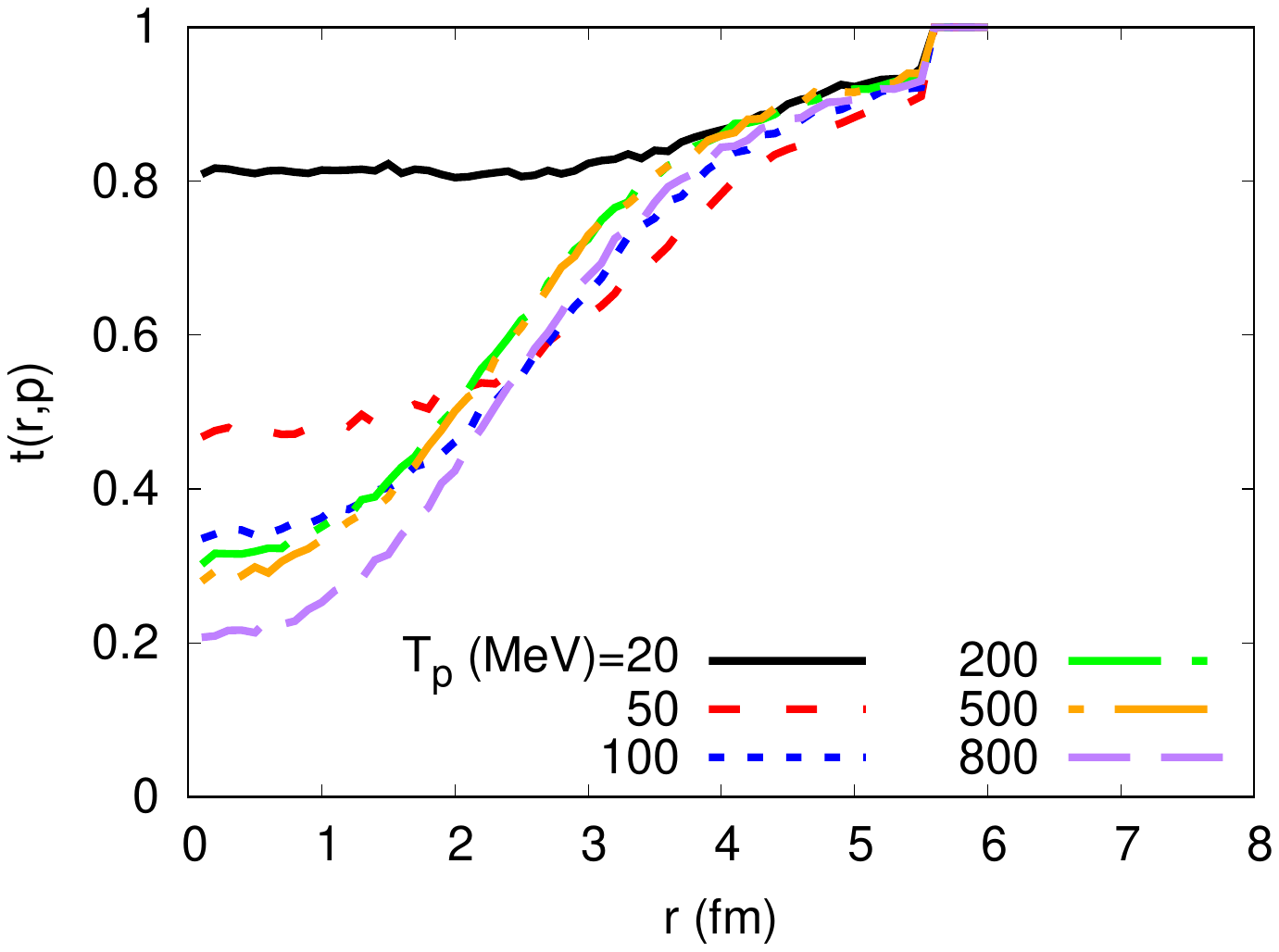}
\hspace{-1cm}
\includegraphics[width=.55\textwidth,angle=0]{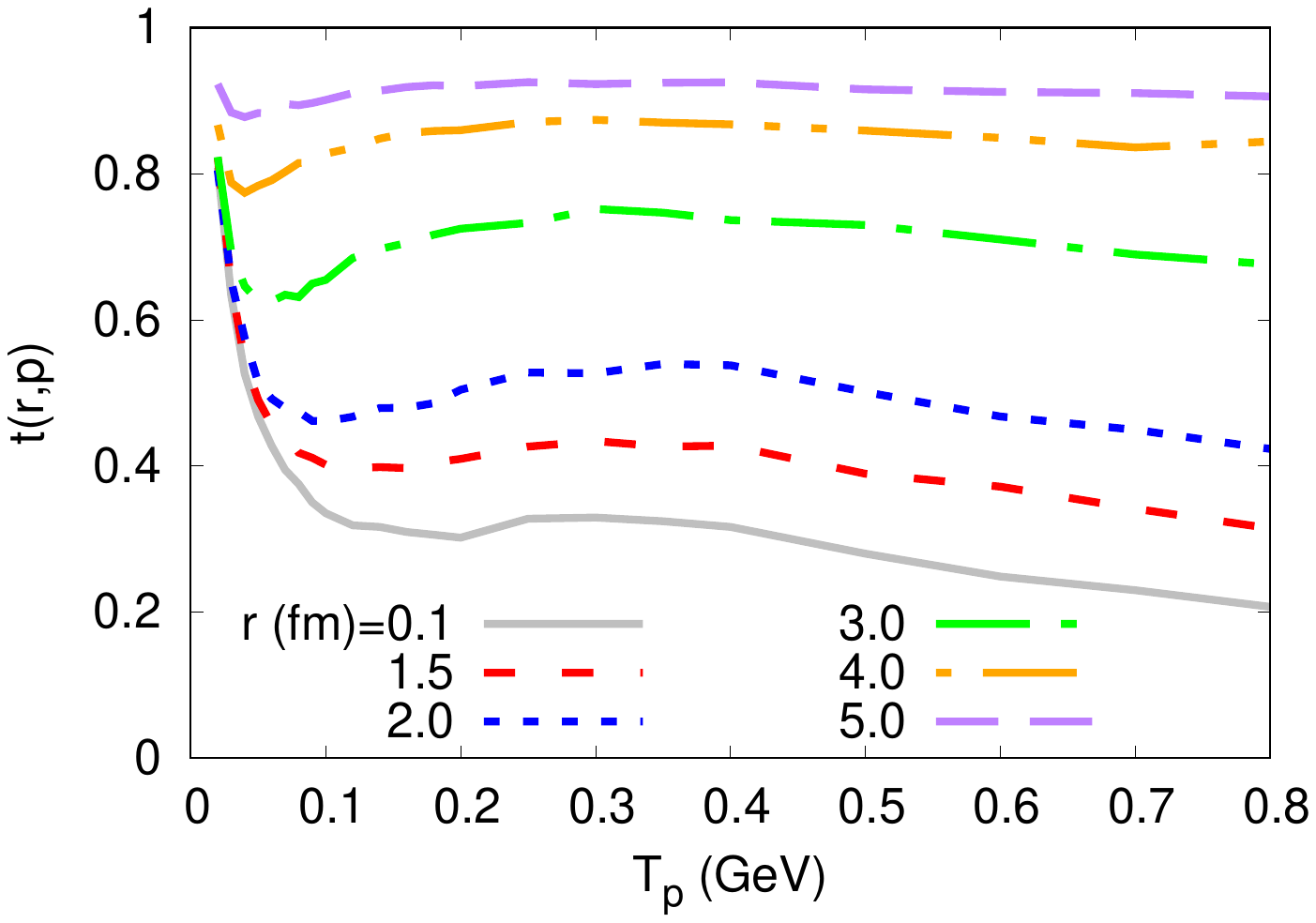}\\
\vspace{-0.5cm}
\hspace{-1cm}
\includegraphics[width=.55\textwidth,angle=0]{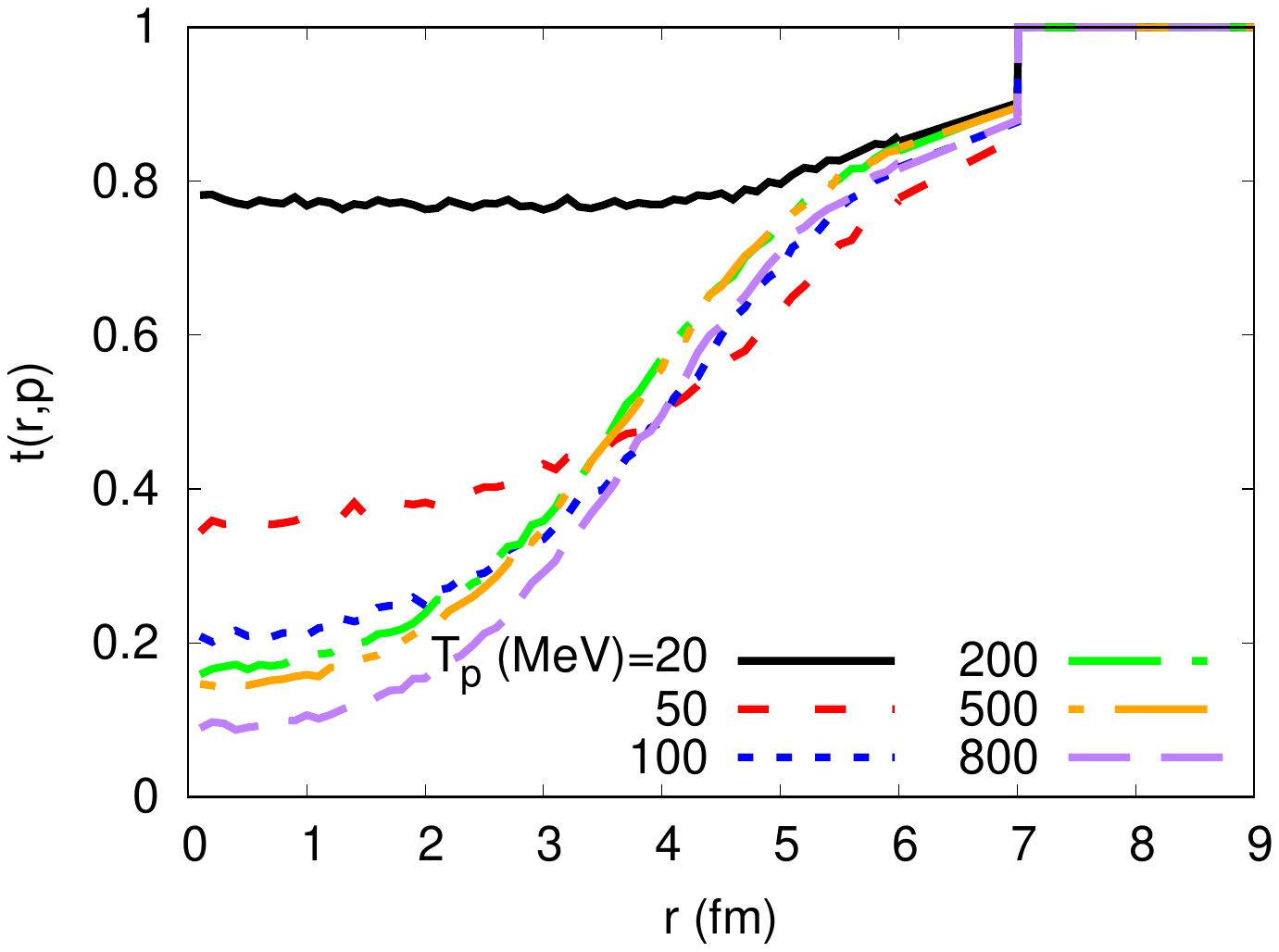}
\hspace{-1cm}
\includegraphics[width=.55\textwidth,angle=0]{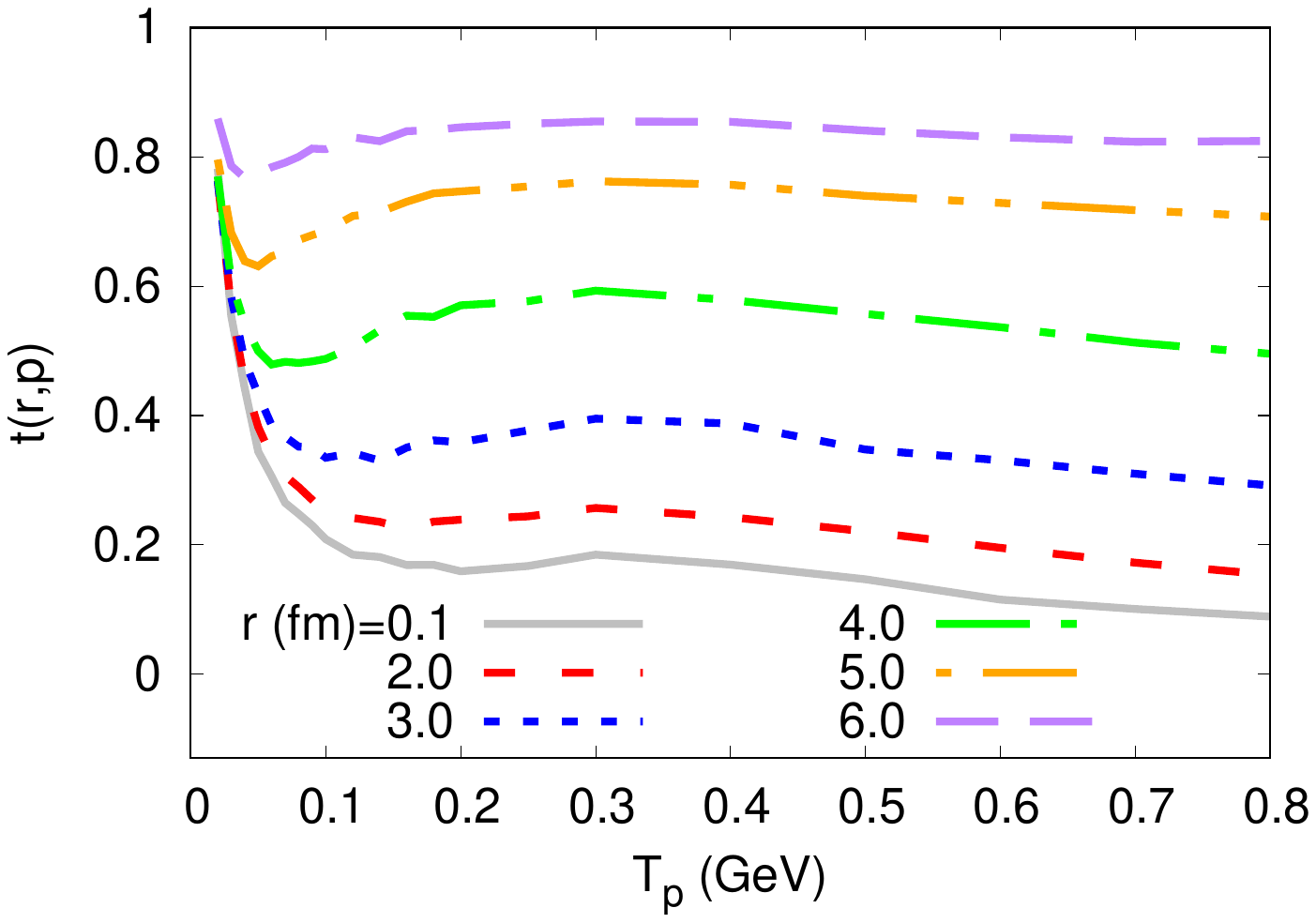}
\vspace{-0.5 cm}
\caption{Local transparency in the NEUT INC, defined as the fraction of protons that do not suffer any inelastic interaction on their way out of the nucleus.
Upper and lower panels are the results for ${}^{12}$C and ${}^{40}$Ca, respectively.  
In the left panels, we plot the transparency as a function of $r$ (proton starting point in the nucleus) for some values of the kinetic energy of the proton. 
In the right panels, we plot the transparency as a function of the kinetic energy of the proton for some $r$ values.
}
\label{fig:NEUT-transparency}
\end{figure*}

\begin{figure*}
\centering
\vspace{-0.5cm}
\hspace{-1cm}
\includegraphics[width=.55\textwidth,angle=0]{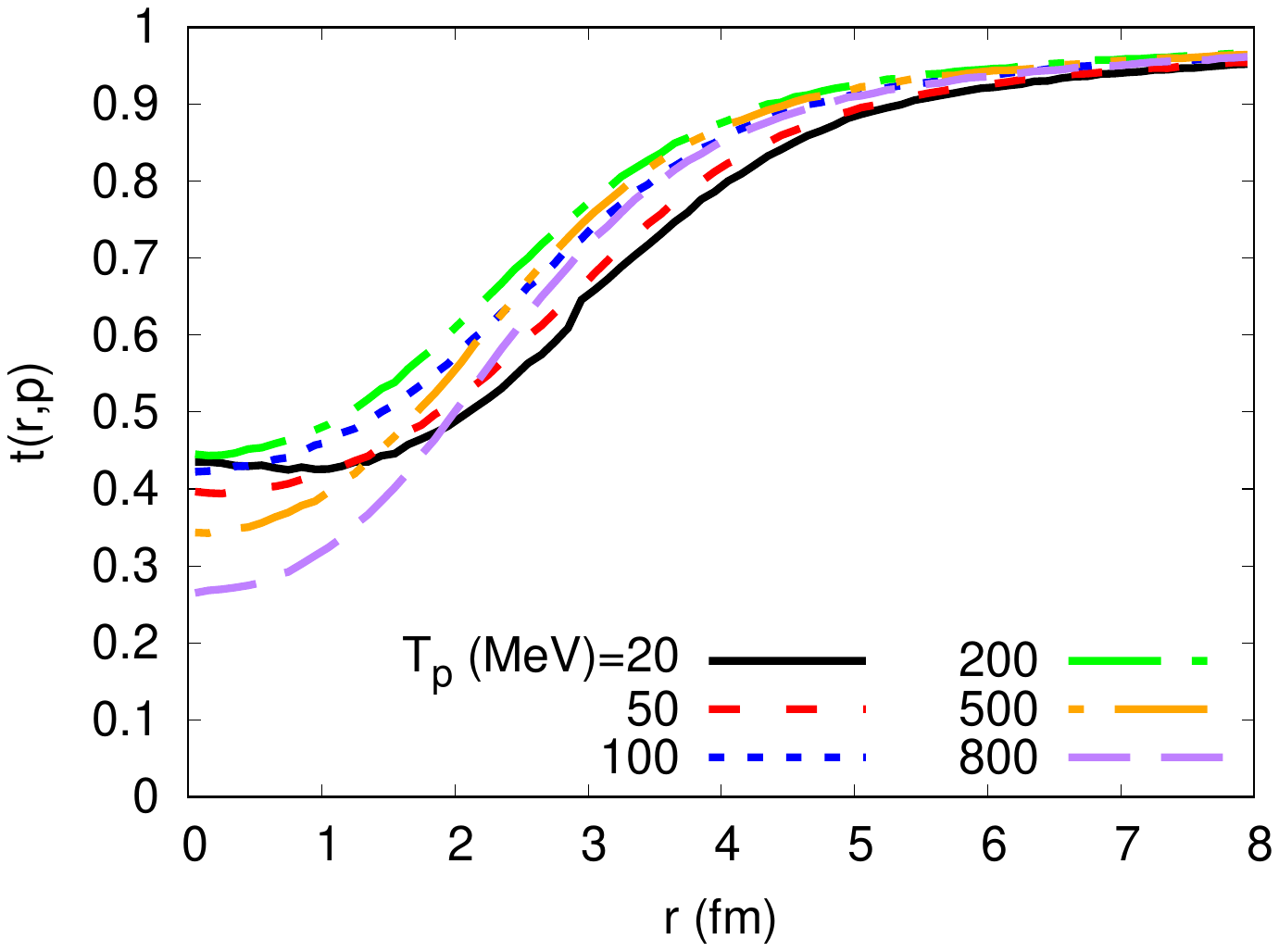}
\hspace{-1cm}
\includegraphics[width=.55\textwidth,angle=0]{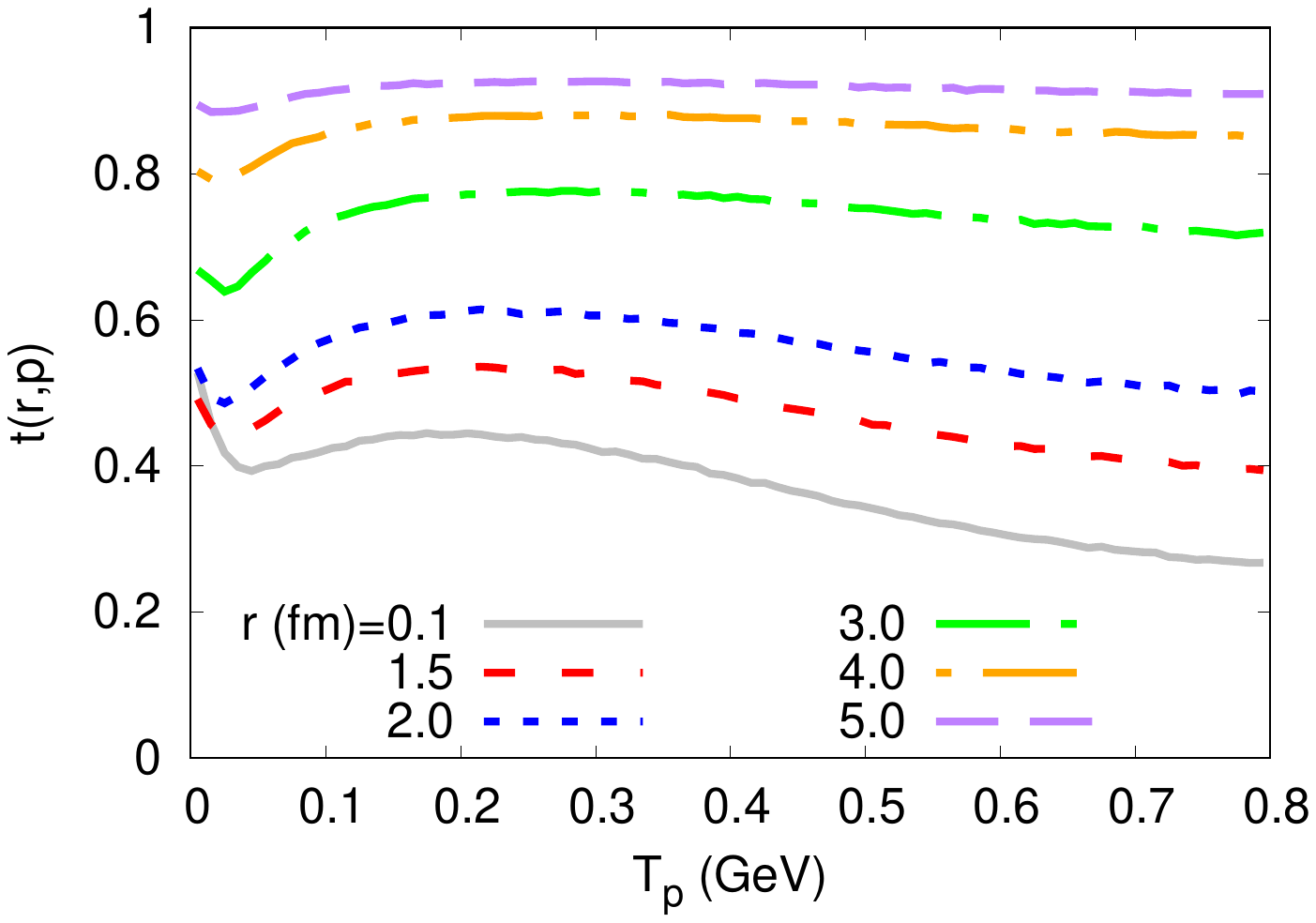}\\
\vspace{-0.5cm}
\hspace{-1cm}
\includegraphics[width=.55\textwidth,angle=0]{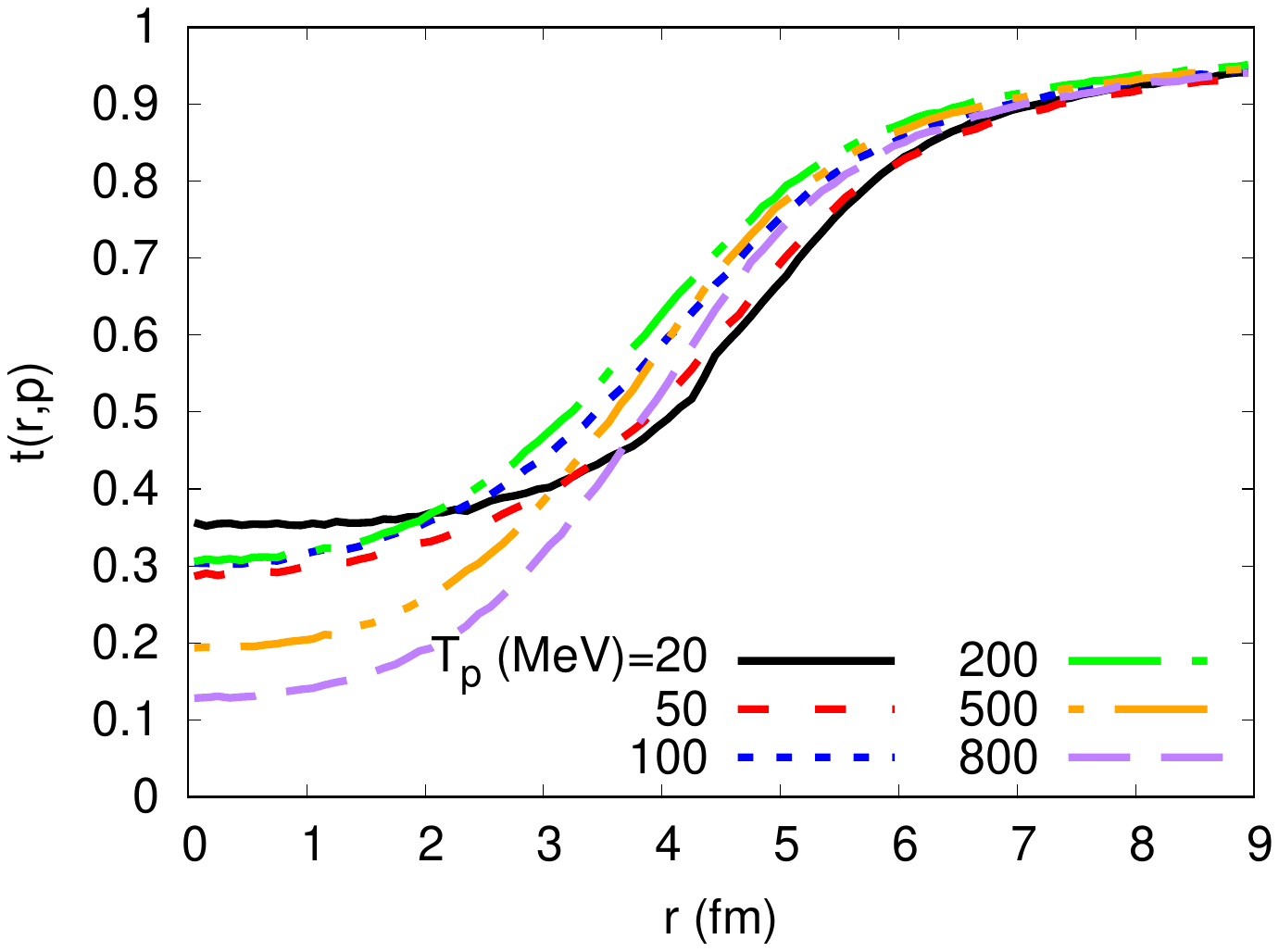}
\hspace{-1cm}
\includegraphics[width=.55\textwidth,angle=0]{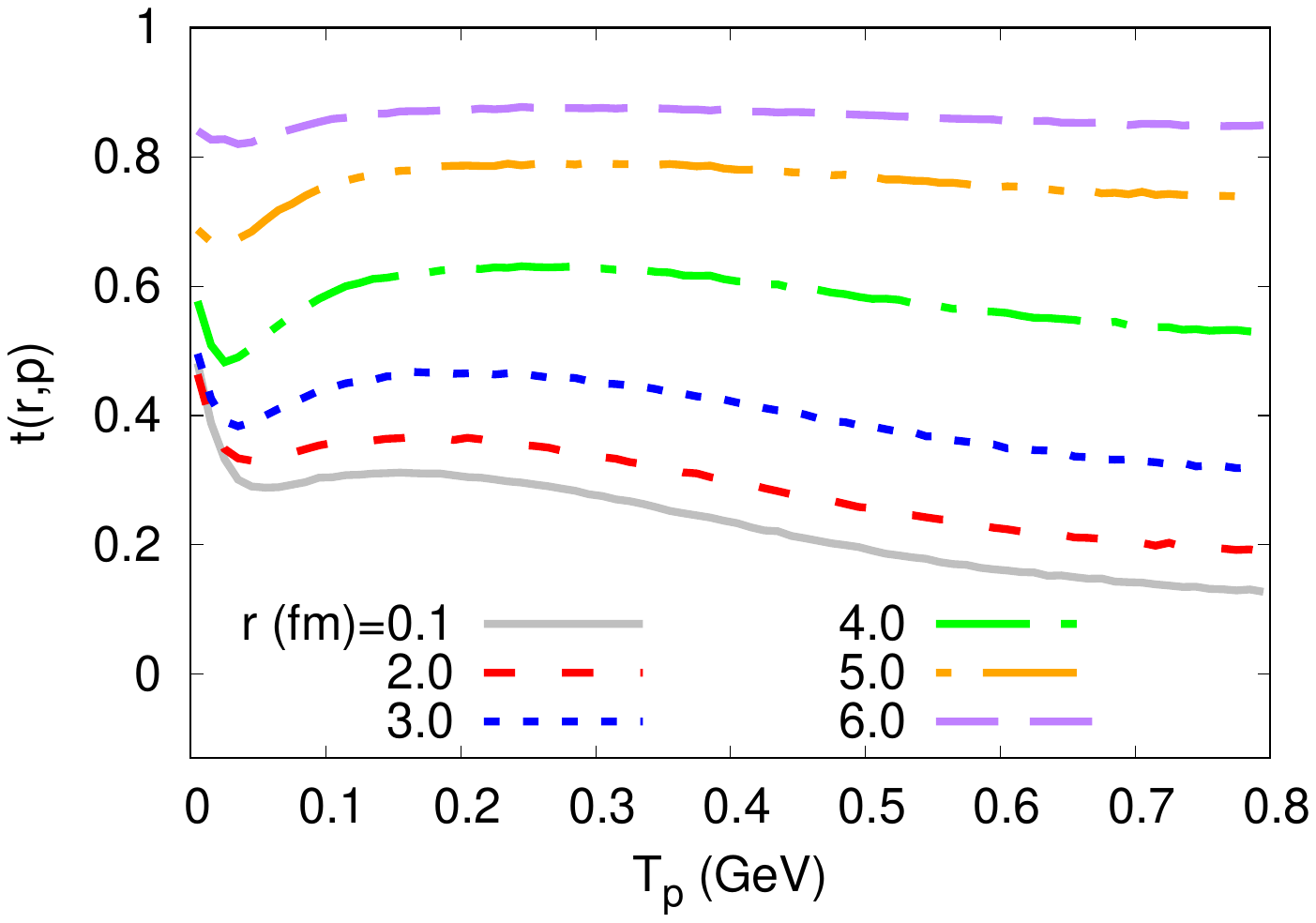}
\vspace{-0.5 cm}
\caption{As Fig.~\ref{fig:NEUT-transparency} but computed with the NuWro INC. 
}
\label{fig:NuWro-transparency}
\end{figure*}

The transparency in the INC model can be readily computed because of its classical nature; in a simulation, a nucleon traveling through the nucleus either interacts or does not.
To define the transparency in the INC, we first define the `local transparency' $t(\vb{r},\vb{p})$ as the probability that a nucleon placed at a point $\vb{r}$ at distance $r$ from the center of the nucleus with momentum $\vb{p}$ escapes without interaction.
Due to spherical symmetry, it is useful to average over all points on a sphere with radius $r$ and define 
\begin{equation}
    t(r,p) \equiv \frac{1}{2}\int \mathrm{d}\cos\theta_{r,p}~ t(\vb{r},\vb{p}), 
\end{equation}
where $\cos\theta_{r,p} = \hat{\vb{r}}\cdot\hat{\vb{p}}$.
This local transparency is readily calculated through simulation in the INC model. 
Results for the NEUT and NuWro INC models are presented in Fig.~\ref{fig:NEUT-transparency}. 
The nuclear transparency is governed by the nucleon mean-free path~\cite{Hayato21,Niewczas19}
\ba 
  \lambda = [\sigma(E)\rho(r)]^{-1}, \label{eq:meanfreepath}
\ea
which depends on the nucleon-nucleon cross section $\sigma(E)$ and the nuclear density $\rho(r)$. 
Aside from its implicit density dependence, the $r$-dependence of $t(r,p)$ is a result of geometry: nucleons placed deeper within the nucleus must traverse a longer path through the medium, thereby increasing their probability of interaction, as illustrated in Figs.~\ref{fig:NEUT-transparency} and \ref{fig:NuWro-transparency}~\footnote{At large $r$, well outside of the nuclear volume, it follows from geometry that $t(r,p) = 1 - \tilde{\sigma}(p)/r^2$, with $\tilde{\sigma}(p)$ proportional to the nucleon-nucleus reaction cross section. This regime is not relevant to our studies and one can cut the calculation at sufficiently large $r$.}.

While $t(r,p)$ is well defined and a crucial quantity in the INC model, it cannot be directly connected to experimental reality.
Hence, typically one considers the nuclear transparency
\begin{equation}
    \label{eq:T_rho_int}
    T(p)[\rho] \equiv \frac{1}{N}\int dr~r^2 \rho(r)~t(r,p),
\end{equation}
where $\rho$ is taken to be the total radial density of the nucleus, and $N$ is its norm.
Hence, the transparency $T(p)[\rho]$ is the probability that a randomly chosen nucleon in the nucleus escapes, when given an initial momentum $p$.
This quantity has served as an important benchmark for INC models~\cite{Niewczas19, Dytman21}.

The notion of transparency is readily generalized by considering a radial distribution function $\rho^{\prime}(r)$ different from the total nuclear density $\rho(r)$ in Eq.~\eqref{eq:T_rho_int}.
That is, we can define a transparency $T(p)\left[ \rho^{\prime} \right]$ for any arbitrary radial distribution function $\rho^{\prime}(r)$. As we show by considering the RDWIA in the next section, this generalization is both theoretically meaningful, and in the two-step approach is directly connected to the cross section for $(e,e^\prime p)$ below the two-nucleon knockout threshold. 

In this work, we use the NEUT INC without the formation zone or Pauli blocking, and include only the nucleon rescattering channel~\cite{Hayato21}.
We use NuWro version 25.11.1~\cite{Golan12} with its default parametrization for carbon. For calcium, we do not include the correlation effects of Ref.~\cite{Niewczas19} and consequently do not apply the strength scaling of Ref.~\cite{Prasad25}.

\subsection{Transparency within RDWIA}

In order to compare with experimental transparency data, usually defined as $\sigma_{\text{data}}/\sigma_\text{PWIA}$, in RDWIA the nuclear transparency is modeled as $\sigma_{\text{ROP}}/\sigma_{\text{PWIA}}$~\cite{Lava04, Ryckebusch05, Kelly05, Martinez06, Butkevich25};
it represents an estimation of the fraction of quasifree one-nucleon knockout (where the final nucleon propagates elastically before leaving the nucleus), relative to the QE cross section at the vertex computed within PWIA. 
However, as discussed previously and illustrated in Fig.~\ref{fig:RL} (Sec.~\ref{subsec:inclusive}), the PWIA~\footnote{PWIA and RPWIA produce essentially identical results for inclusive and exclusive cross sections, so we do not make a distinction between them in this work; for an in depth discussion about relativistic effects in RPWIA we refer to~\cite{Caballero98a}.} is insufficient to accurately represent the QE cross section at the vertex. Consequently, interpreting this quantity as `the number of nucleons that do not rescatter' is conceptually flawed.

An alternative definition, which could better resemble the concept of nuclear transparency, consists in using in the denominator of the ratio a model that provides a better description of the elementary vertex (i.e., of the inclusive cross section in the QE peak). 
In our RDWIA framework, it would be $\sigma_{\text{data}}/\sigma_{\text{r(ROP)}}$ or  $\sigma_{\text{data}}/\sigma_{\text{EDRMF}}$.
Then, nuclear transparency would be modeled as $\sigma_{\text{ROP}}/\sigma_{\text{r(ROP)}}$ or $\sigma_{\text{ROP}}/\sigma_{\text{EDRMF}}$. 

When defined through such a ratio, nuclear transparency generally depends on the specific kinematics under consideration, namely, the lepton and nucleon variables. Also, because the shapes of the cross sections in the numerator and denominator do not match, the classical assumption of factorized inelastic final-state interactions breaks down.
For this measure to be useful, the cross sections should be integrated over a sufficiently large region of phase space.
For this purpose, we define the ratio as
\begin{flalign}
 T_\text{model} = \frac{\de\sigma_\text{ROP}/\de T_p}{\de\sigma_\text{model}/\de T_p}\,,\label{eq:T_RDWIA}
\end{flalign}
the numerator is the single-differential cross section with the ROP model, while the denominator is the same quantity computed with a `real-potential' model, in this work r(ROP) or EDRMF, or the zero-potential RPWIA. 
The electromagnetic cross section, contrary to the weak one, is not well defined in the full phase space due to the divergence at $Q^2\rightarrow0$ (or, equivalently, at zero lepton scattering angle). Therefore, we define $d\sigma/dT_p$ in a restricted kinematic range:
\ba 
\frac{\de \sigma}{\de T_p} = \int_{q_{min}}^{q_{max}} {\de q\, \frac{\de\sigma}{\de T_p \de q}} = \int_{-1}^{+1} {\de\psi\, \frac{\de\sigma}{\de T_p \de\psi}}\,,
\ea 
where $q_{min}$ and $q_{max}$ are the minimum and maximum momentum transfer for QE scattering within the relativistic Fermi gas (RFG) model~\cite{Foundations17}. In the RFG, the QE peak is fully contained in the region $-1\leq\psi\leq1$, where $\psi=\psi(q,\omega)$ is the QE scaling variable, defined e.g. in Eq.~(44) in~\cite{Amaro20}. 
For our interest, for a given energy transfer $\omega$~\footnote{Considering the mean-field levels as energy eigenstates, the missing energy $E_m$ is a fixed value for each level; then, for a given level and $T_p$ value, one gets $\omega$ from energy conservation: 
\ba 
\omega \approx T_p + E_m\,,
\ea
where the nuclear recoil has been neglected.}, $q_{min}$ corresponds to $\psi=-1$ and $q_{max}$ to $\psi=+1$. 
Defined in this manner, the transparency depends exclusively on the proton kinetic energy, exhibiting only a mild sensitivity to the initial electron energy and the cutoff values $q_{min}$ and $q_{max}$~\footnote{We have verified that one finds similar results for the cross section ratios computed for neutrino scattering using the code developed for Ref.~\cite{Nikolakopoulos24}, where the integral is over the full phasespace of the MicroBooNE experiment.}.
Consequently, it satisfies the criteria for a robust definition; a dependence on additional kinematic variables (considered later in this work) would otherwise confound the (classical) concept of transparency discussed above.

\begin{figure*}
\centering
\vspace{-0.7cm}
\hspace{-0.7cm}
\includegraphics[width=.53\textwidth,angle=0]{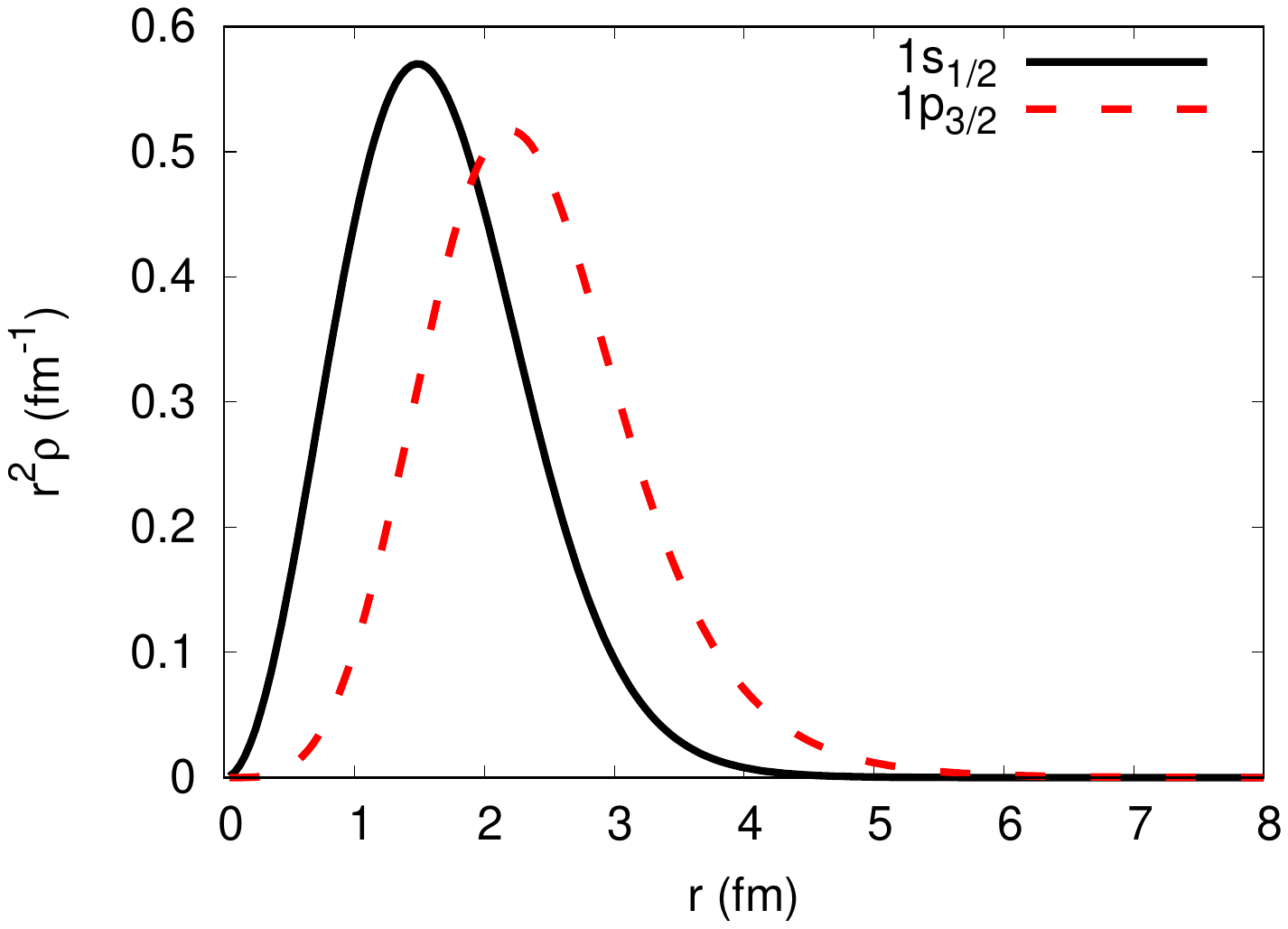}
\hspace{-0.7cm}
\includegraphics[width=.53\textwidth,angle=0]{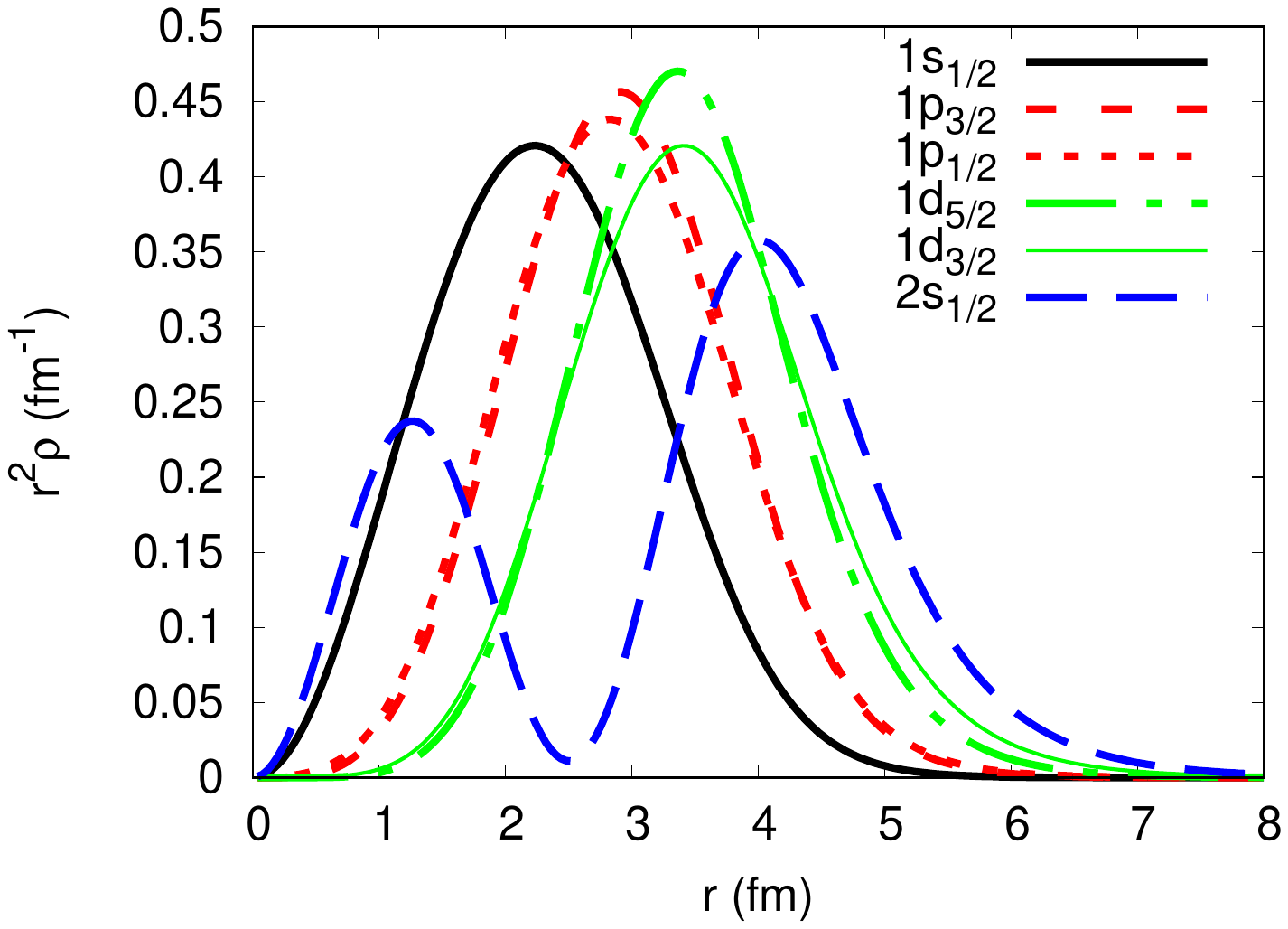}\\
\vspace{-0.5cm}
\hspace{-0.7cm}
\includegraphics[width=.53\textwidth,angle=0]{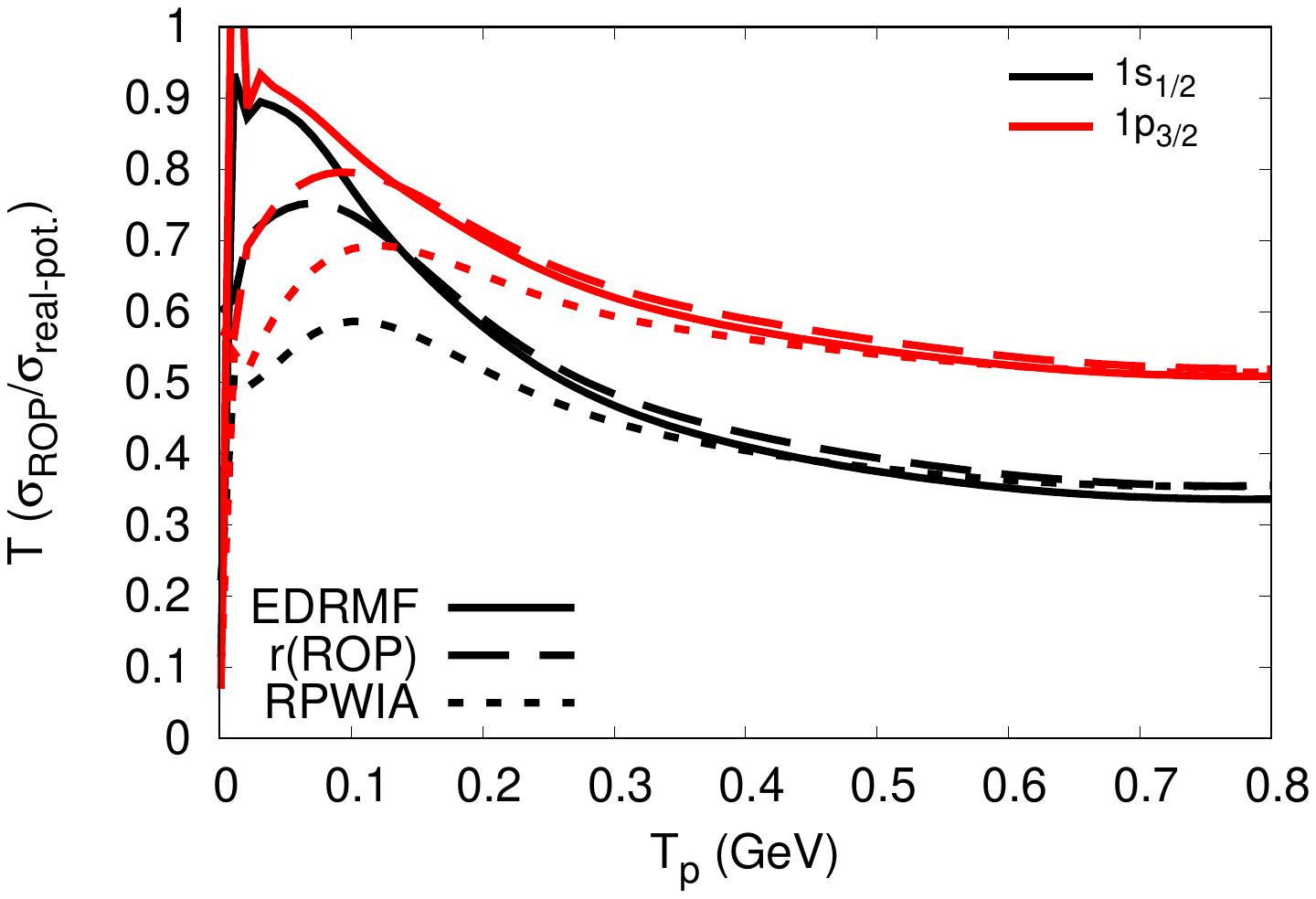}
\hspace{-0.7cm}
\includegraphics[width=.53\textwidth,angle=0]{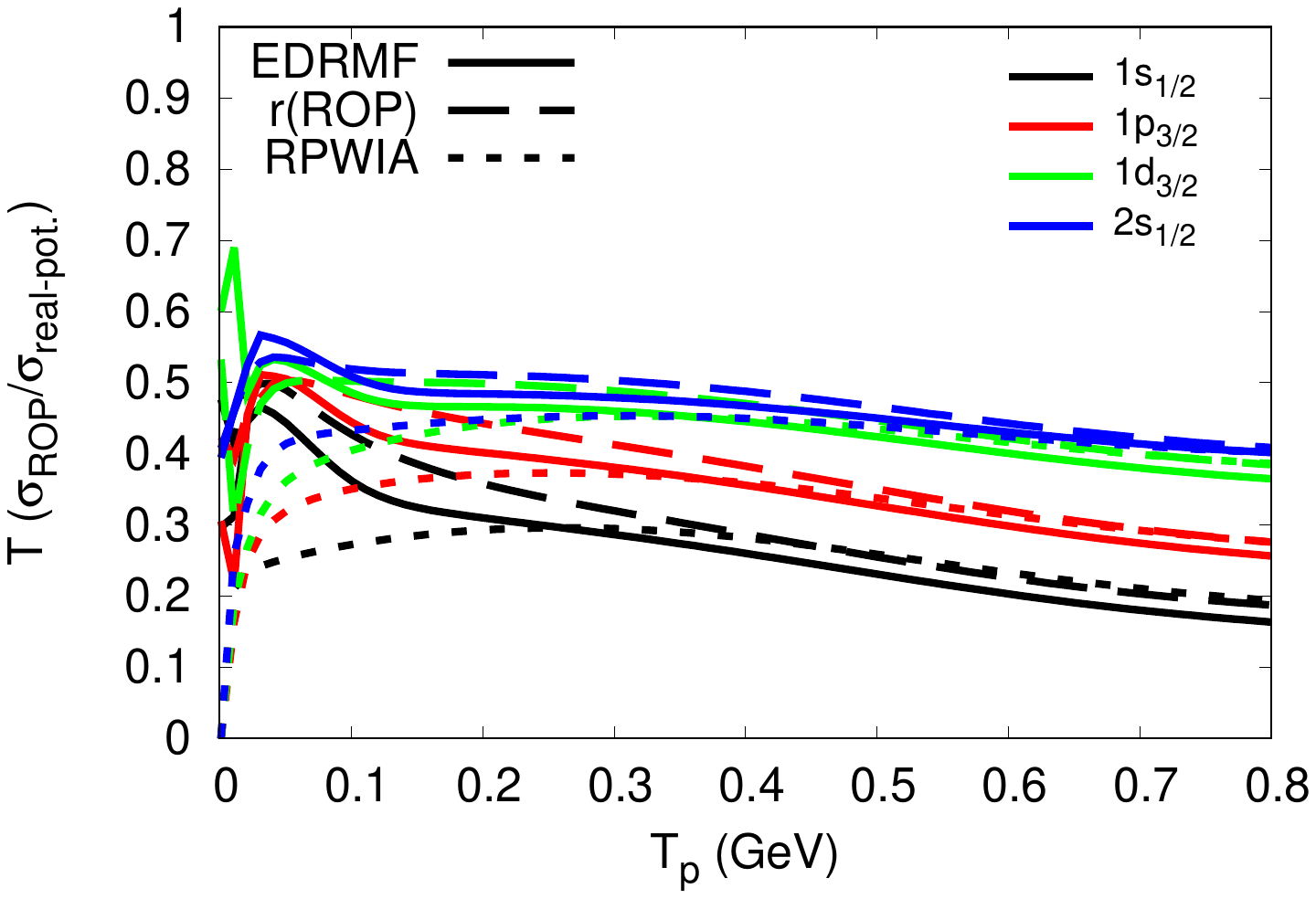} \\
\vspace{-1cm}
\hspace{-0.7cm}
\includegraphics[width=.53\textwidth,angle=0]{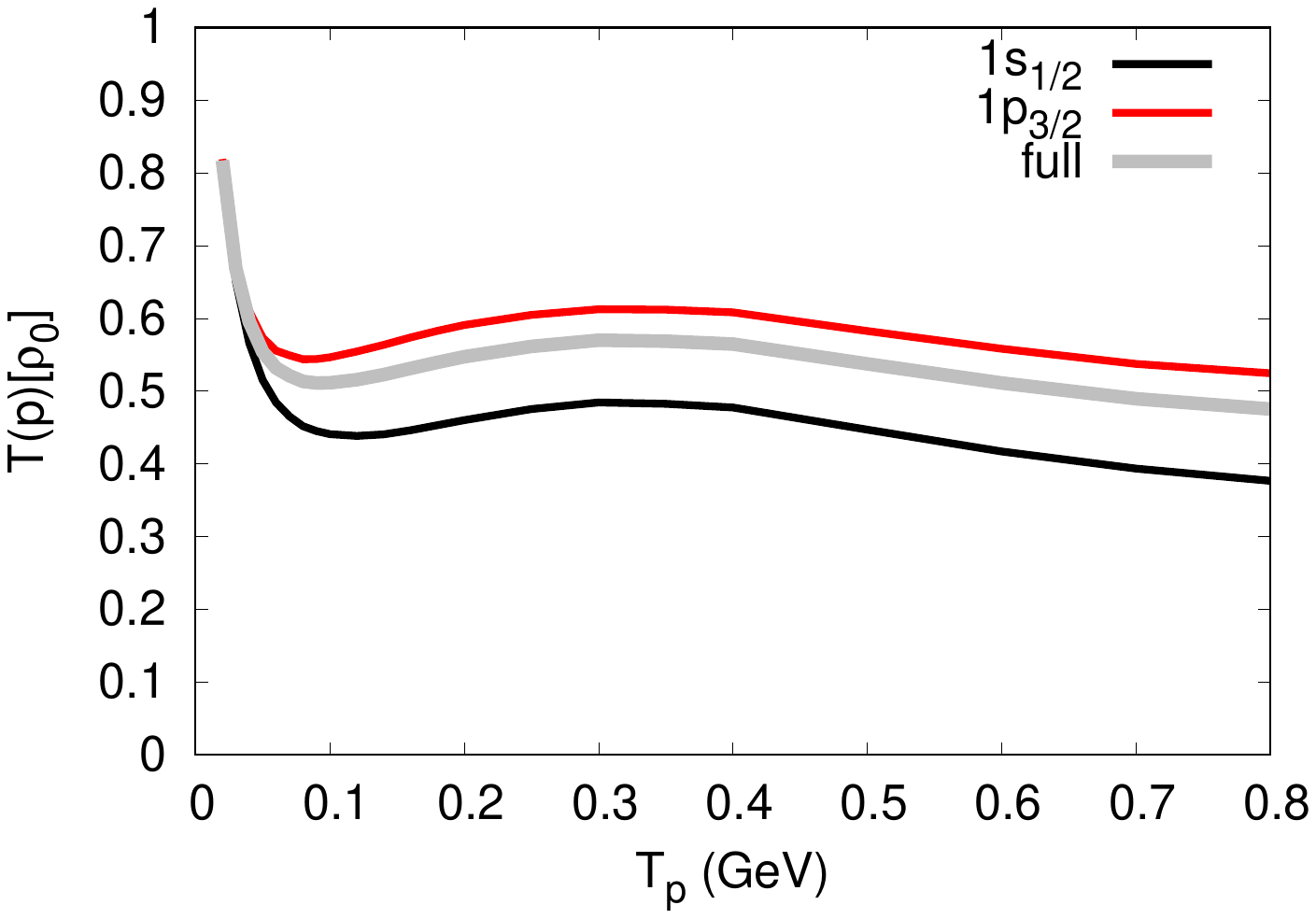}
\hspace{-0.7cm}
\includegraphics[width=.53\textwidth,angle=0]{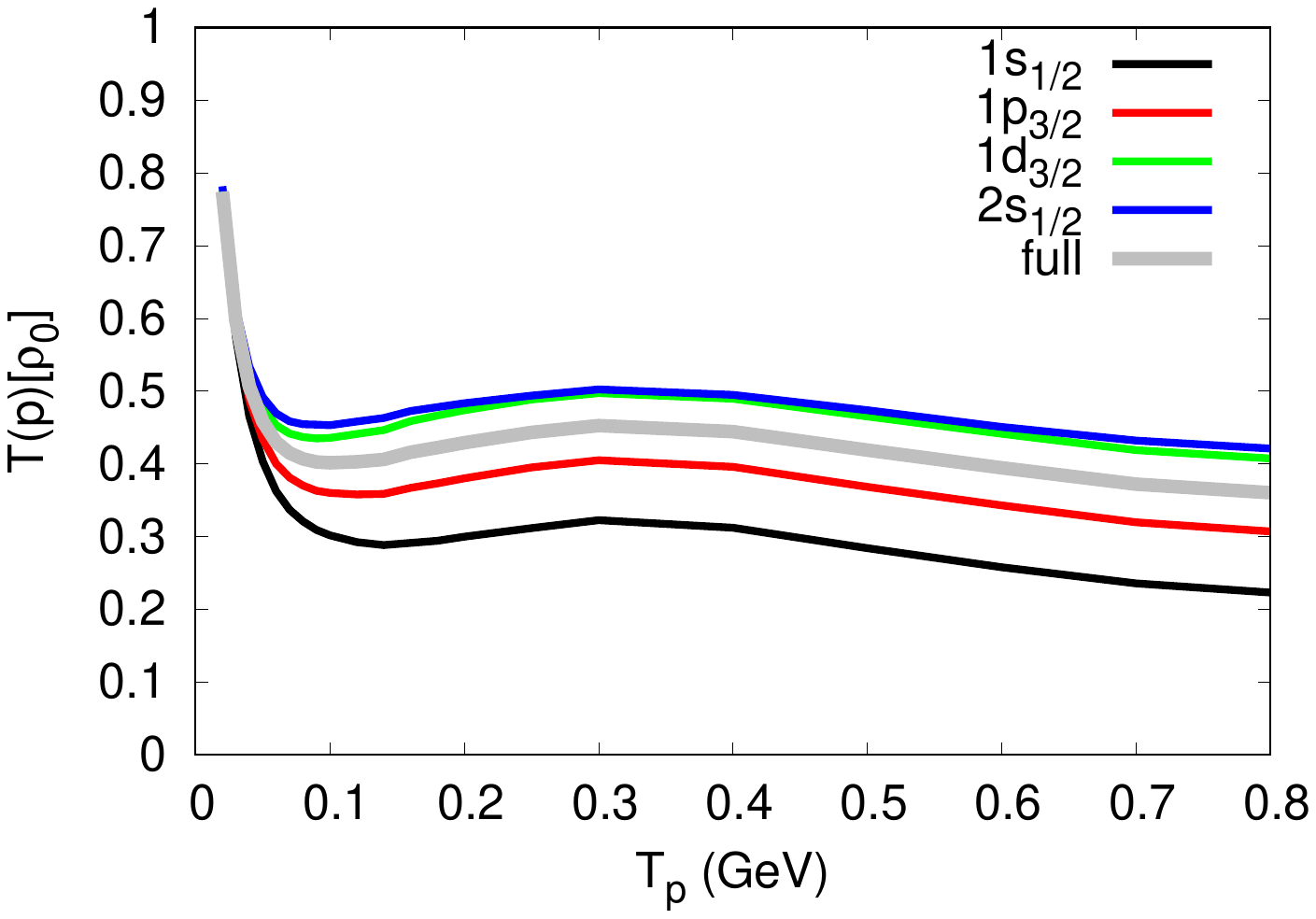}\\
\vspace{-1cm}
\hspace{-0.7cm}
\includegraphics[width=.53\textwidth,angle=0]{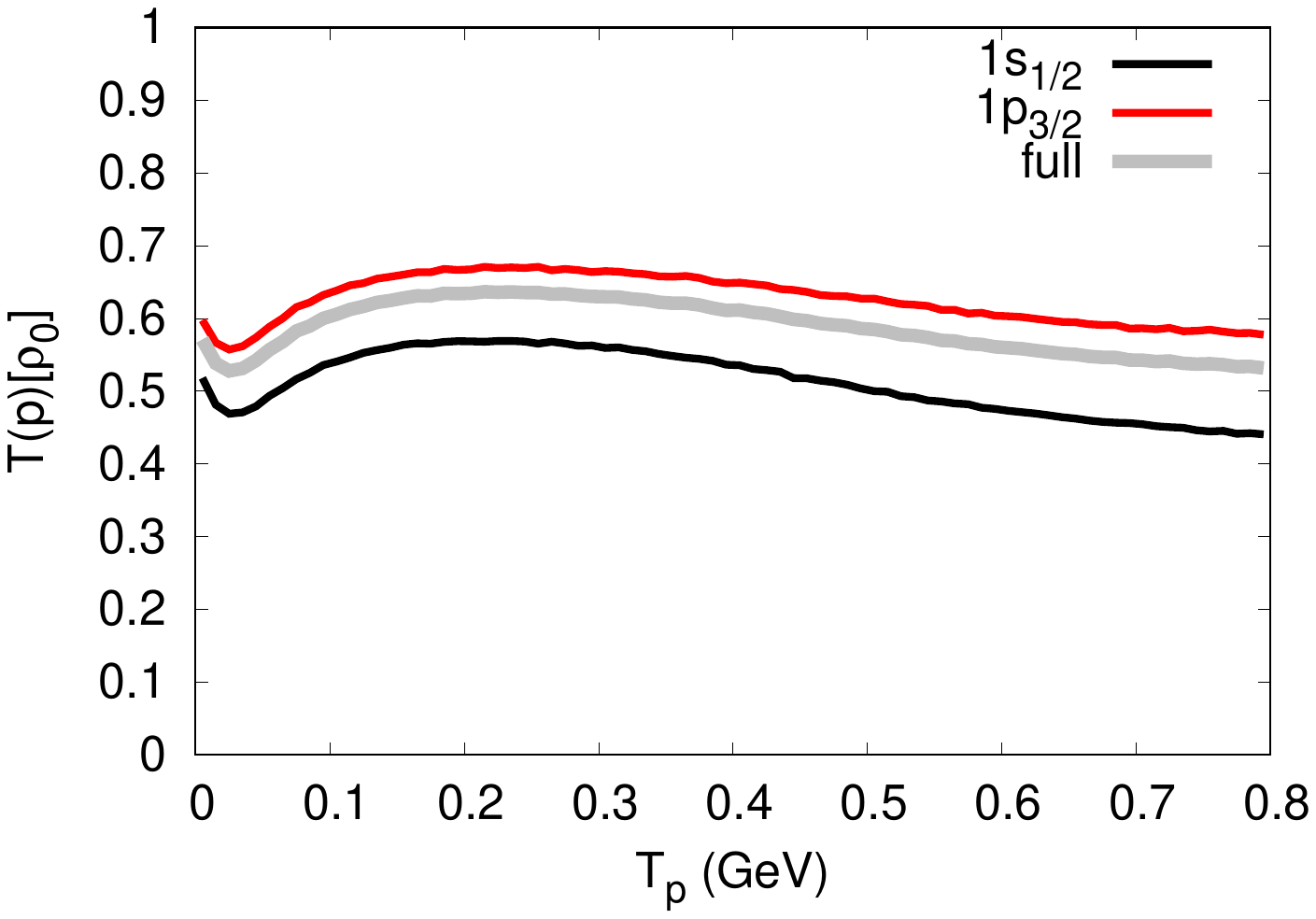}
\hspace{-0.7cm}
\includegraphics[width=.53\textwidth,angle=0]{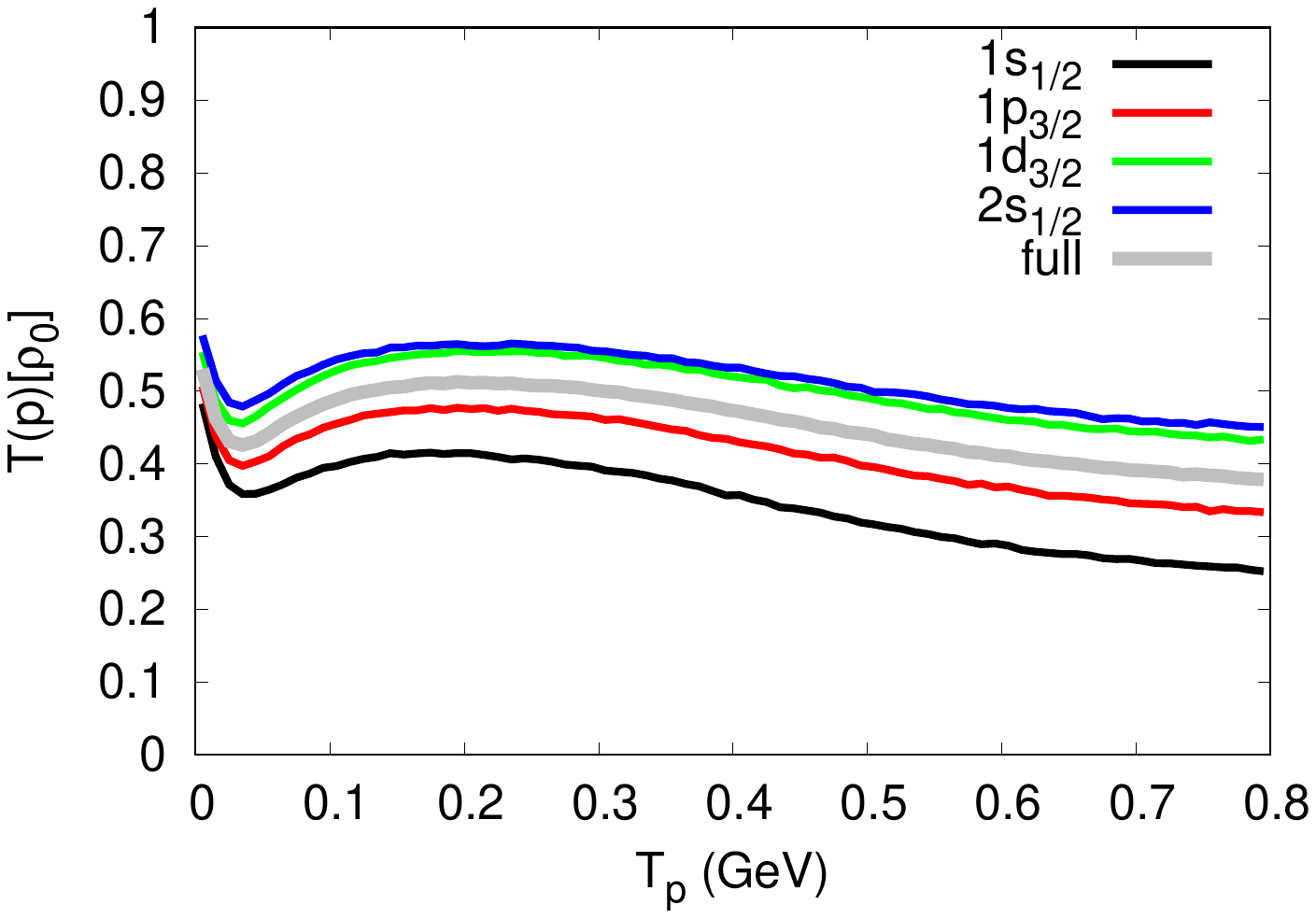}
\vspace{-0.7cm}
\caption{Results for ${}^{12}$C and ${}^{40}$Ca are on the left and right panels, respectively. 
(First row) Density for each level in ${}^{12}$C and ${}^{40}$Ca, all of them normalized to 1 and computed with the RMF model of Refs.~\cite{NLSH,Horowitz81}. The factor $r^2$ is included.
(Second row) Transparency for different levels and with three RDWIA models (see text). Each level is represented with a different color, while different models correspond to different type of line.
To simplify the plot, the results for the $1d_{5/2}$ and $1p_{1/2}$ levels in ${}^{40}$Ca are not shown, but they are very similar in shape and magnitude to the $1d_{3/2}$ and $1p_{3/2}$ levels, respectively.
(Third row) The results obtained with the NEUT INC using Eq.~\eqref{eq:T_rho_int}. The thicker gray line is the full NEUT transparency, i.e., the result of Eq.~\ref{eq:T_rho_int} when using the full RMF nuclear density for $\rho_0(r)$. 
(Fourth row) As third row but for the NuWro INC model.
\label{fig:density_transp}
}
\end{figure*}

\begin{table}[h!]
\centering
\caption{Average position $\langle r\rangle$ (fm) for the different RMF levels in ${}^{40}\text{Ca}$ and ${}^{12}\text{C}$ shown in the upper panels in Fig.~\ref{fig:density_transp}.}
\label{tab:mean-radius}
\setlength{\tabcolsep}{5.5pt}
\begin{tabular}{lcccccc}
\hline\hline
 & \textbf{$1s_{1/2}$} & \textbf{$1p_{3/2}$} & \textbf{$1p_{1/2}$} & \textbf{$1d_{5/2}$} & \textbf{$1d_{3/2}$} & \textbf{$2s_{1/2}$} \\ \hline
${}^{40}\text{Ca}$ & 2.33 & 2.95 & 2.92 & 3.45 & 3.52 & 3.39 \\
${}^{12}\text{C}$  & 1.66 & 2.40 & ---  & ---  & ---  & ---  \\ \hline\hline
\end{tabular}
\end{table}

The analysis of $(e,e'p)$ exclusive data, studied in Section~\ref{sec:Results}, indicates that this ratio, defined in Eq.~\eqref{eq:T_RDWIA}, depends on the density of the level from which the nucleon originated. 
This effect is predicted by the RDWIA~\cite{Lava04, Ryckebusch05, Nikolakopoulos24, Butkevich25} and is shown in Fig.~\ref{fig:density_transp}.
In the upper panels in Fig.~\ref{fig:density_transp}, we show the radial distribution of the different single-particle states $r^2\rho_\kappa(r)$ as a function of $r$.
The left panel corresponds to ${}^{12}$C and the right panel to ${}^{40}$Ca, both calculated with the RMF model~\cite{Horowitz81,NLSH}.
The second row shows the transparency defined by Eq.~\eqref{eq:T_RDWIA}.

Irrespective of the model assumed for the denominator, $T_\text{model}$ exhibits a clear shell dependence. 
Generally, outer shells, which are characterized by a larger mean radius $\langle r\rangle$, display higher transparency. The mean radii are listed in Table~\ref{tab:mean-radius}. 
This trend holds for all states except the $2s_{1/2}$ orbital, which exhibits the highest transparency despite not having the largest $\langle r \rangle$, demonstrating that the shape of the spatial distribution plays a role.

Given the clear classical analogy that nucleons that originate from larger radii should have a lower probability to interact with the nucleus, it is possible to capture this effect with the INC.
This is shown in the third and fourth rows of Fig.~\ref{fig:density_transp}.
Here we compute the transparency obtained for the different shells by averaging the local transparency $t(r,p)$ obtained in the NEUT and NuWro INC models using the RMF radial density of each single-particle state. 
That is, we show $T(p)[\rho_\kappa]$ as defined in Eq.~\eqref{eq:T_rho_int}. 
Above 50-100 MeV, the INC captures the shell dependence obtained in the RDWIA approach well, thereby confirming the classical interpretation.

The level-dependent transparency could be straightforwardly implemented in an event generator by placing the struck nucleon in the nucleus according to the density profile of the level from which it originated, namely $\rho_\kappa(r)$~\footnote{According to the results of Fig.~\ref{fig:density_transp} and Table~\ref{tab:mean-radius}, it would be a good approximation to place the nucleon simply using the mean radius $\langle r\rangle$ instead of the shell density.}.
Obviously, this is possible provided the underlying nuclear model incorporates shell structure; it cannot be accounted for by infinite nuclear-matter models. 
Although now most event generators can use shell-model approaches, such as PWIA with realistic spectral functions~\cite{Hayato21,Ershova22,Isaacson23,Betancourt23} or RDWIA~\cite{McKean25,Franco-Munoz26}, to our knowledge, they do not account for this specific effect. 
Indeed, they utilize the \textit{full} nuclear density to decide where to place the nucleon before its propagation.

This shell-effect has been studied in~\cite{Nikolakopoulos22}, but only cross sections integrated over a large region of phase space were considered. However, this effect is essential for improving data comparisons in scenarios sensitive to the underlying shell structure. 
A clear example are the exclusive $(e,e'p)$ data discussed in the next section, which should serve to constrain and refine both the elementary-vertex and INC models. 
Beyond electron scattering, this shell effect may also play a critical role in modeling neutrino-nucleus cross sections. For instance, the recent dataset from~\cite{Marzec25}, obtained using a  KDAR~\footnote{Kaon decay at rest (KDAR).} mono-energetic neutrino beam, exhibits a clear sensitivity to these shell effects. 
This dataset was recently analyzed in~\cite{McKean26,Franco-Patino26} within a framework that combines the RDWIA approach and the NEUT INC, but without taking into account the shell dependence of the transparency.

Importantly, this level-dependent effect and its integration into INC models are not limited to the QE process; it also applies to final-state hadrons in pion production, multi-nucleon knockout processes and other reaction mechanisms.

\section{Analysis of electron scattering data}\label{sec:Results}

\subsection{Inclusive longitudinal response}\label{subsec:results-inclusive}

We compare several RDWIA predictions with inclusive electron scattering data to benchmark the performance of the various models, a critical step in identifying the most suitable elementary-vertex formulation for MC event generators. 
We stress that the inclusive cross section is blind to the INC model, being completely determined by the elementary-vertex model.
To avoid complications from MEC effects, which, for moderate momentum transfer $q$, predominantly manifest in the transverse response~\cite{Amaro10a}, we focus on the longitudinal response.

\begin{figure*}
\centering
\hspace{-0.8cm}
\includegraphics[width=.52\textwidth,angle=0]{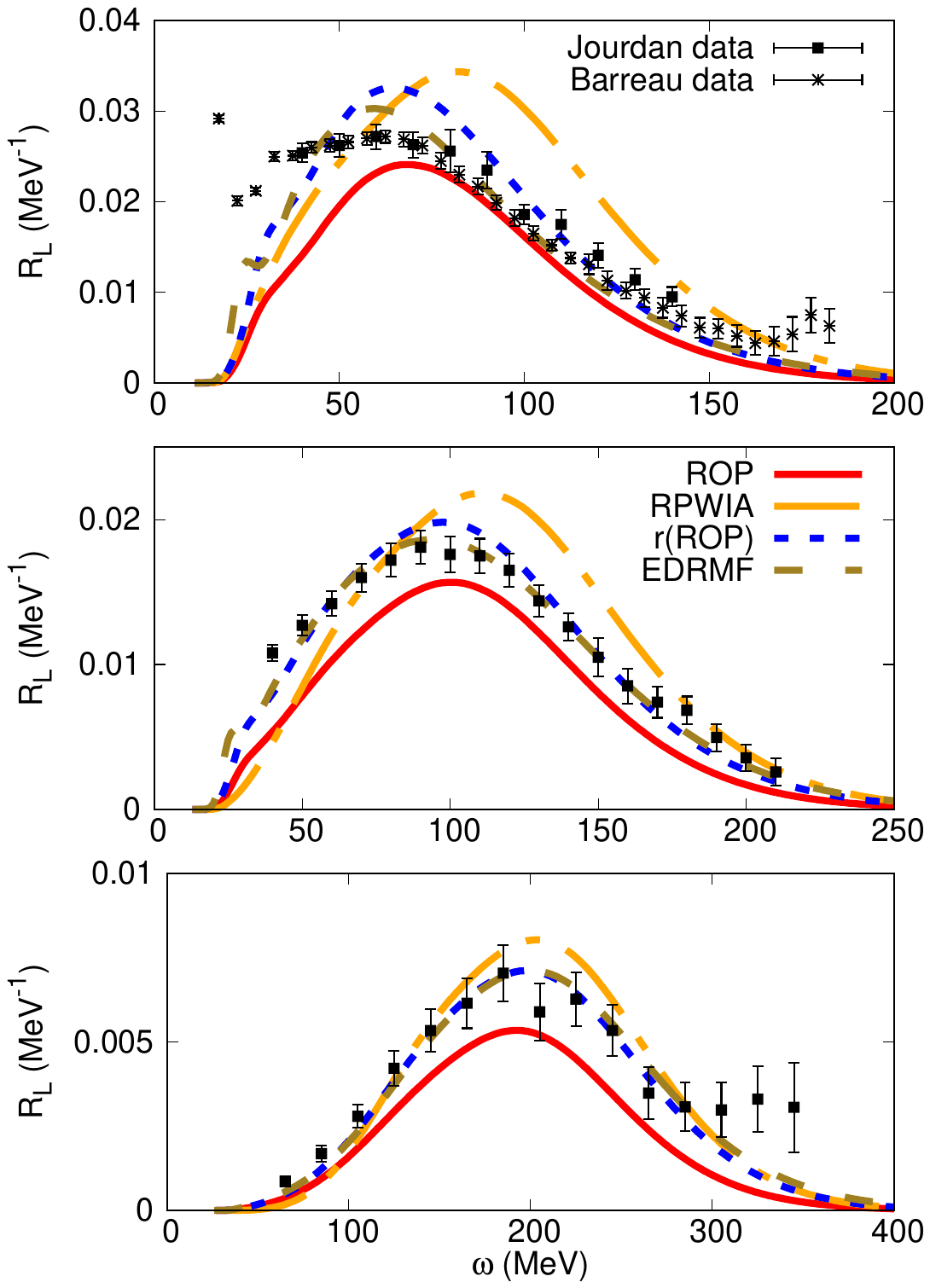}
\hspace{-1cm}
\includegraphics[width=.52\textwidth,angle=0]{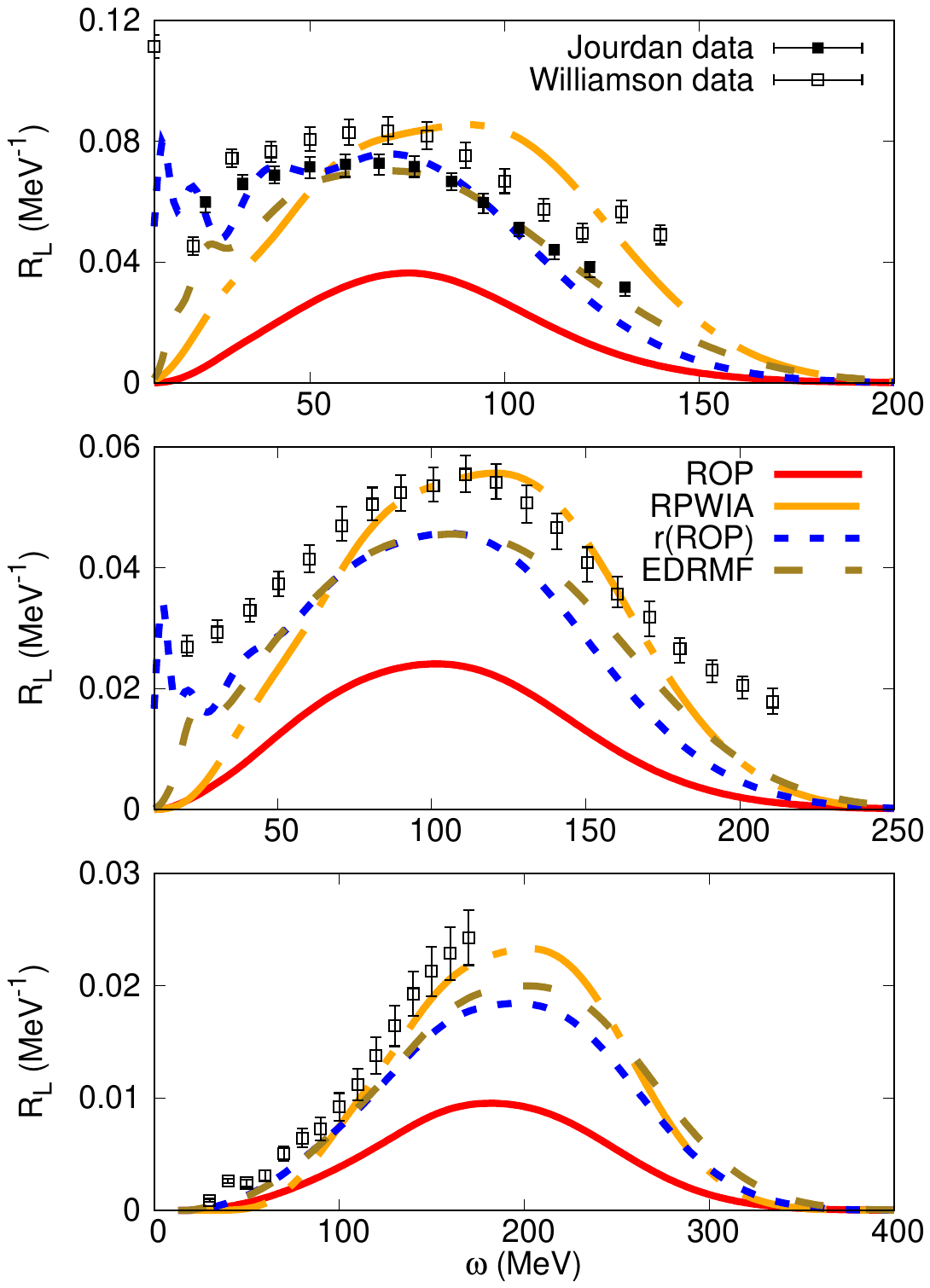}
\vspace{-0.8cm}
\caption{EDRMF, r(ROP), RPWIA and ROP predictions compared to longitudinal electromagnetic response ${}^{12}$C data from~\cite{Barreau83,Jourdan96} (left panels) and ${}^{40}$Ca data from~\cite{Jourdan96,Williamson97,Meziani84,Meziani85} (right panels). Each panel corresponds to a fixed $q$: 300, 380 and 570 MeV (left panels); and 300, 380 and 550 MeV (right panels). }
\label{fig:RL}
\end{figure*}

In the left panels in Fig.~\ref{fig:RL}, we show ${}^{12}$C$(e,e')$ longitudinal response data, for three different values of the momentum transfer $q$ and the predictions from the EDRMF, r(ROP) and ROP. For the optical potentials we use the energy-dependent A-independent ${}^{12}$C (EDAI-C) potential of~\cite{Cooper09}.
In the right panels, we show analogous results but for ${}^{40}$Ca$(e,e')$. In this case, the complex optical potential is the EDAI-Ca~\cite{Cooper09}.

The EDRMF gives a satisfactory description of the inclusive longitudinal response for carbon. 
The r(ROP) is very close to EDRMF except at the lowest $q$ and $\omega$ values (top panels), where due to non-orthogonality issues~\cite{Gonzalez-Jimenez20}, gives a slightly worse description of data. 
The ROP, i.e. the cross section for the quasifree one-nucleon knockout process, has a similar shape, but its magnitude is only a fraction of the full strength. 
RPWIA is shifted with respect to the previous models and data, overestimating them in the peak. Only at the highest $q$ (lower panels), the shape and magnitude of the RPWIA gets closer to data. 

A similar discussion applies in the case of calcium results, but in this case, the EDRMF and r(ROP) models underestimate the medium and largest $q$ data. This is discussed below.

The spectroscopic factors used in the inclusive results here are as given in Ref.~\cite{Franco-Munoz25} for carbon and calcium.
However, the results shown here, contrary to those in~\cite{Franco-Munoz25}, do not include the contribution attributed to scattering of short-range correlated (SRC) pairs. 
In the spectral function~\cite{Benhar08}, the SRC contribution appears as a continuum at high $E_m$ and $p_m$, and in the inclusive scattering results, it contributes to the high $\omega$ tail of the spectrum.
To clarify, our current calculations are restricted to one-nucleon knockout from the mean-field part of the spectral function. Based on the spectroscopic factors employed in this work~\cite{Franco-Munoz25}, this corresponds to 10.2 nucleons (85\%) in ${}^{12}$C 
and 27.1 nucleons (67.5\%) in ${}^{40}$Ca, 
assuming symmetric proton and neutron distributions for both nuclei. 
This relatively low nucleon count in calcium is the reason of the underestimation of the inclusive results. 
The modeling of SRC pairs is complex and beyond the scope of this work, therefore,  to maintain a focused discussion, we omit their contribution from our present predictions.

Another feature is that the quasifree one-nucleon knockout contribution to the inclusive cross section, namely the ROP curve, is significantly lower in calcium than in carbon. 
This effect originates from the broader nuclear density of the former: a nucleon traversing a larger nucleus exhibits, on average, a higher probability of undergoing inelastic interactions before leaving the system.
This was anticipated by the results of Fig.~\ref{fig:density_transp}, which show that on average carbon is more transparent than calcium.

\subsection{Analysis of $(e,e'p)$ results}\label{subsec:results-exclusive}

\begin{figure*}
\centering
\hspace{-1cm}
\includegraphics[width=.45\textwidth,angle=0]{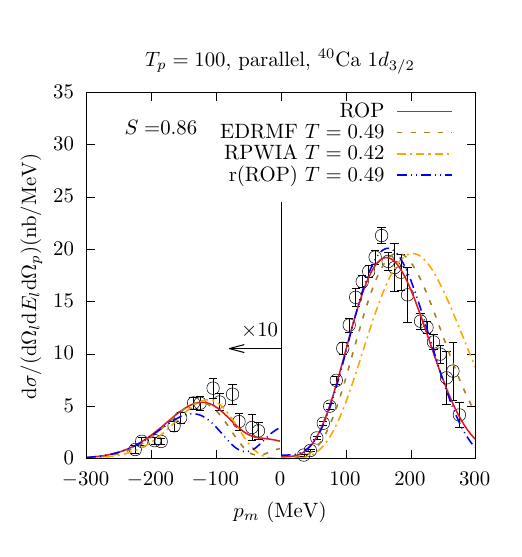}
\hspace{-1cm}
\includegraphics[width=.45\textwidth,angle=0]{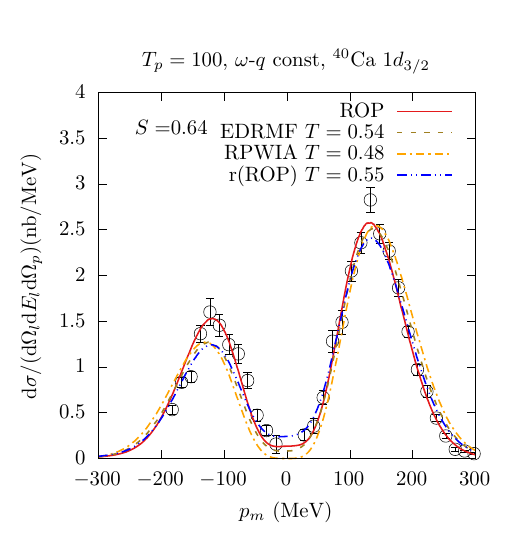}\\
\vspace{-1cm}
\hspace{-1cm}
\includegraphics[width=.45\textwidth,angle=0]{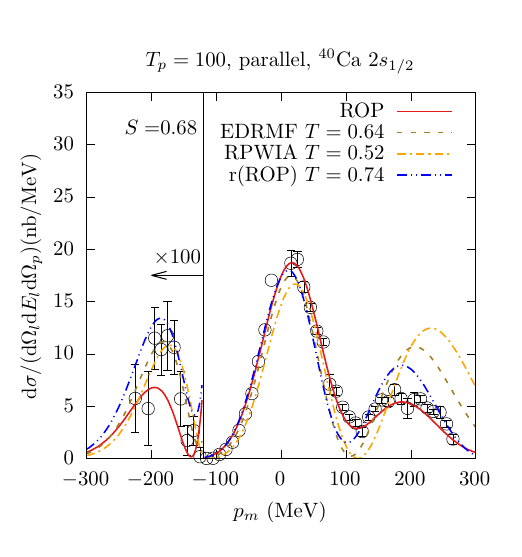}
\hspace{-1cm}
\includegraphics[width=.45\textwidth,angle=0]{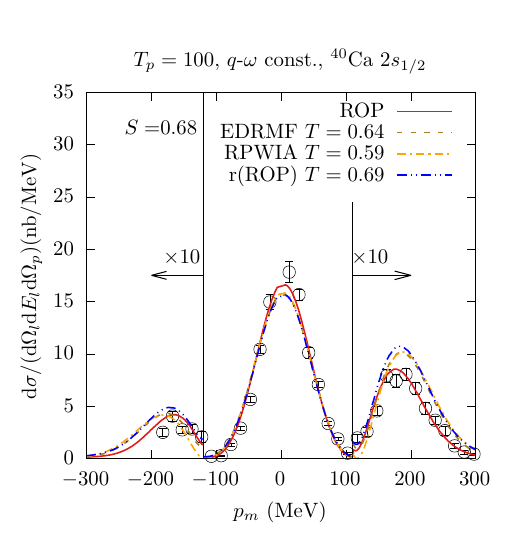}
\hspace{-0.5cm}
\vspace{-0.5cm}
\caption{Model predictions compared to ${}^{40}\text{Ca}(e,e'p){}^{39}\text{K}$ data from NIKHEF~\cite{Kramer89,Kramer-PhD}. First and second row are $1d_{3/2}$ and $2s_{1/2}$, respectively. First and second column are parallel and $q-\omega$ constant kinematics, respectively. 
We used $E_i=483.2$ MeV in all cases, and $q=450$ MeV for the $q-\omega$ constant kinematics. 
The four panels correspond to $T_p=100$ MeV. 
Model predictions are multiplied by the spectroscopic factor (SF) indicated in the up-left corner in each panel. 
The predictions of the RPWIA, r(ROP) and EDRMF models are furthermore multiplied by the transparency factors obtained from Eq.~\eqref{eq:T_excl}, indicated in the key and summarized in Table~\ref{tab:transparencies}.
Note that the values shown in Table~\ref{tab:transparencies} are the average $T^\text{exp-PS}$ value between the $q-\omega$ const. and parallel kinematics.
}
\label{fig:exclusive_40Ca}
\end{figure*}

\begin{figure*}
\centering
\includegraphics[width=.4\textwidth,angle=0]{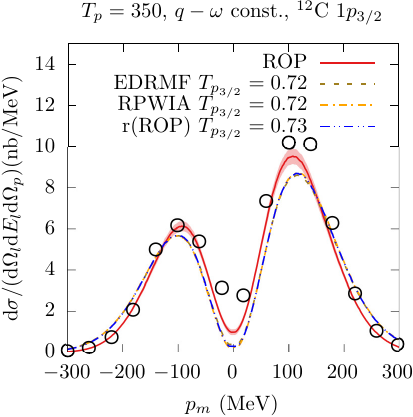}
\includegraphics[width=.4\textwidth,angle=0]{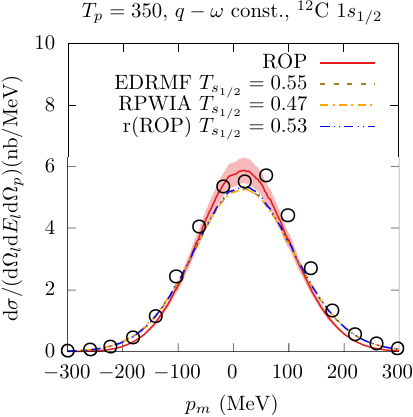}\\

\includegraphics[width=.4\textwidth,angle=0]{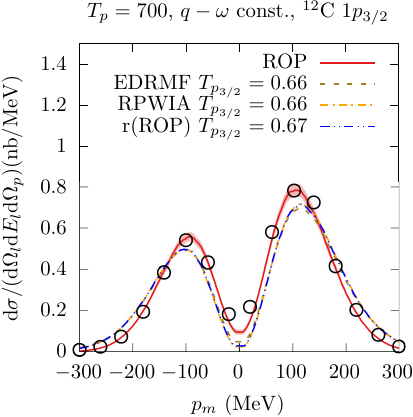}
\includegraphics[width=.4\textwidth,angle=0]{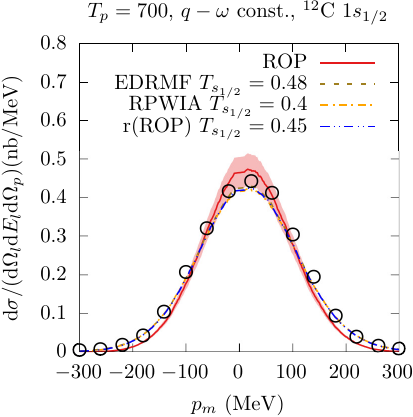}\\

\includegraphics[width=.4\textwidth,angle=0]{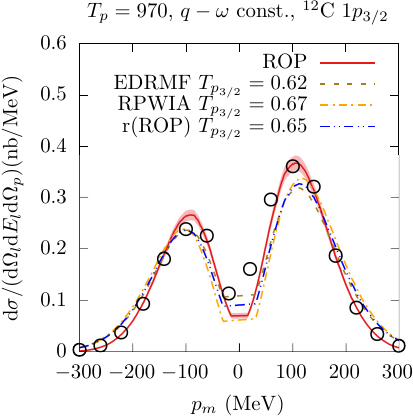}
\includegraphics[width=.4\textwidth,angle=0]{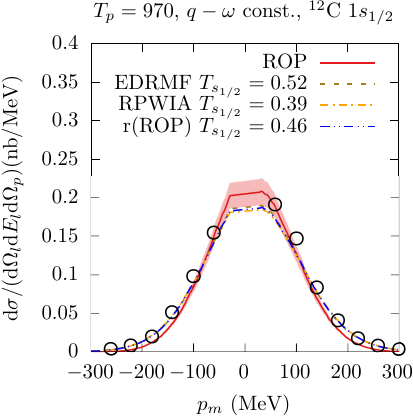}

\caption{Model predictions compared to ${}^{12}$C$(e,e'p)$ JLab data at $q$-$\omega$ constant~\cite{Dutta03}. 
First, second and third row correspond to $Q^2=0.6$, $1.2$ and $1.8$ GeV$^2$, and $T_p=350$, $700$ and $970$ MeV, respectively. 
We used $E_i=2445$ MeV for the kinematics of the first and second row, and 3245 MeV for the kinematics of the third row. 
Left panel are data with $10<E_m<25$ MeV, dominated by $1p_{3/2}$; right panel are data with $30<E_m<50$ MeV, dominated by the $1s_{1/2}$.
The normalization of the cross sections is obtained from the empirical spectral function and given by Eqs.~\eqref{eq:nrm_E1}-\eqref{eq:nrm_Gaus}.
The predictions of the RPWIA, r(ROP) and EDRMF models are furthermore multiplied by the transparency factors obtained from Eq.~\eqref{eq:T_excl}, indicated in the key and summarized in Table~\ref{tab:transparencies}.}
\label{fig:exclusive_12C}
\end{figure*}

Figures~\ref{fig:exclusive_40Ca} and \ref{fig:exclusive_12C} show comparisons of ROP with different $(e,e'p)$ results: ${}^{40}$Ca$(e,e'p)^{39}$K~\cite{Kramer89} and ${}^{12}$C$(e,e'p)$~\cite{Dutta03}.
Experimental data~\cite{Dutta03,Kramer89}, as well as most of the published theoretical analyses, were originally published as reduced cross sections and presented on logarithmic scale.
In this work, we wanted to highlight the regions where the cross section is large rather than focusing on the shape and tails of the distributions; therefore, we have converted the experimental data to absolute cross sections using the usual de Forest's off-shell cross section~\cite{Forest83} for the kinematics detailed in the caption of Figs.~\ref{fig:exclusive_40Ca} and \ref{fig:exclusive_12C}, and present the results in linear scale.

Exclusive $(e,e'p)$ data, where a nucleon is knocked out from a level at missing energy $E_\kappa$ below the two-nucleon knockout threshold can be described using a complex optical potential model. 
In our approach, the ROP prediction, as in the inclusive case, is multiplied by a constant spectroscopic factor $S_\kappa$, which depends on the level.

The $2s_{1/2}$ and $1d_{3/2}$ hole states in the ${}^{40}$Ca data studied here are identified as narrow peaks in the missing energy spectrum~\cite{Kramer89}, and both lie below the two-nucleon knockout threshold.
In this case, we choose the spectroscopic factor such that the ROP model fits the magnitude of the experimental data~\footnote{The primary objective of this work is not to extract spectroscopic factors from analysis of $(e,e'p)$ data, but rather to discuss the underlying methodology. Detailed and systematic statistical analyses of this type have been comprehensively addressed in previous works (see e.g. Refs.~\cite{Udias93,Kelly05,Giusti11,Atkinson18}).}. 
However, one should be careful in the interpretation of these SFs, in particular, if the goal is to extract spectroscopic factors to be used in the modeling of the inclusive cross section. 
States are often fragmented over a relatively broad $E_m$ range and there are intruder states (those that do not appear in the IPSM); this has been shown for different nuclei by experiments with excellent energy resolution, e.g.~\cite{Van-Der-Steenhoven88, Morrow05, Mougey76, Kramer89}.

The situation for the carbon data is somewhat different~\cite{Dutta03}. In this case, the data for $1p_{3/2}$ and $1s_{1/2}$ are obtained in larger windows of missing energy $\Delta E_1 = [10,25]$ and $ \Delta E_2 = [30,50]$ MeV, respectively; the latter above the two-nucleon knockout threshold ($27.19$~MeV in ${}^{12}$C ~\cite{Wang21}).
The $1s_{1/2}$ strength is smeared out over a broad range of missing energy, and can contribute to the $\Delta E_1$ data.
In this case, we used the same parametrization of the missing energy profile as in the NEUT implementation of the RDWIA to determine the normalization~\cite{McKean25,Franco-Munoz26}.
The $1p_{3/2}$ hole state is completely contained in $\Delta E_1$ and does not contribute to the higher energy window. The cross sections shown in the figures are obtained as
\begin{align}    
    \label{eq:nrm_E1}
    \de\sigma_{\Delta E_1} &= S_{p_{3/2}} \de\sigma_{p_{3/2}} + N_{\Delta E_1} \de\sigma_{s_{1/2}},\\
    \label{eq:nrm_E2}
    \de\sigma_{\Delta E_2} &= N_{\Delta E_2} ~\de\sigma_{s_{1/2}}.
\end{align}
The contribution of the $1s_{1/2}$ cross section to an energy window with width $\Delta E$ is computed from the integral of the Gaussian parametrization of the broad $1s_{1/2}$ hole state
\begin{align}
    N_{\Delta E} &= \frac{S_{s_{1/2}}}{\sigma_{s_{1/2}} \sqrt{2\pi}} \times \nonumber \\ &\int_{\Delta E} \de E~ \exp[-\frac{1}{2}\left( \frac{E-E_{s_{1/2}}}{\sigma_{{s_{1/2}}}}\right)^2]\,.
    \label{eq:nrm_Gaus}
\end{align}
The spectroscopic factors $S_\kappa$, central value and width of the Gaussian are given in Appendix A of Ref.~\cite{McKean25}.
The uncertainties shown in Fig.~\ref{fig:exclusive_12C} are obtained by varying both $S_\kappa$ by $4\%$, $E_{s_{1/2}}$ by $1\%$ and the width $\sigma_{s_{1/2}}$ by $1$ MeV. One obtains $N_{\Delta E_1} = 0.055 \pm 0.016$ and $N_{\Delta E_2} = 0.65 \pm 0.05$.

Were our simple approach for incorporating constant spectroscopic factors valid, then the spectroscopic factor extracted from two data sets, corresponding to similar missing energy windows, should be the same (within uncertainties) for a given orbital, regardless of the rest of kinematical conditions. 
This works relatively well for the JLab datasets shown in Fig.~\ref{fig:exclusive_12C} and for the two NIKHEF datasets corresponding to the $2s_{1/2}$ level, Fig.~\ref{fig:exclusive_40Ca}. 
However, a unique spectroscopic factor value does not allow us to fit the two $1d_{3/2}$ NIKHEF datasets, we found notably different values.
This discrepancy may stem from differences in the normalization of the experimental data, arising for instance from minor variations in the selected missing-energy windows, or from nuclear effects beyond our model. Notably, long-range correlations, which are not explicitly included in our framework, quench the cross section at relatively low momentum transfers~\cite{Jachowicz21,Gonzalez-Jimenez20}.

An important question arises regarding the predictive capability of MC event generators when describing these $(e,e'p)$ data. 
Recently, in Ref.~\cite{Butkevich25}, the neutrino event generator GENIE was compared with various ${}^{16}\text{O}(e,e'p)$ datasets using different elementary-vertex models, specifically, several Fermi-gas-based approaches and an `effective spectral function' model. The primary conclusion was that none of these elementary-vertex models, when combined with various GENIE INC models, could describe the data well.
In this work, we address a different aspect of this problem. If the RDWIA (with real potentials) is utilized as the elementary-vertex model, as done in~\cite{Nikolakopoulos22,Nikolakopoulos24,McKean25,Franco-Patino26,McKean26,Franco-Munoz26}, what role does the INC play? 

For the NIKHEF experiment (Fig.~\ref{fig:exclusive_40Ca}), the narrow spectrometer acceptances and the specific kinematic selection, tailored to maximize the quasifree single-nucleon knockout cross section, significantly simplify the impact of the INC. Essentially, it filters out events in which the struck nucleon undergoes inelastic interactions. 
Indeed, if a secondary interaction in the INC occurs, the missing energy increases and the event does not contribute to the data.
Consequently, the final prediction reduces to the elementary-vertex cross section scaled by a constant transparency factor predicted by the INC~\footnote{This is not the case if the INC includes effects where the nucleon can exchange momentum with the system without losing energy, in which case the INC can affect the shape of the cross section, e.g. if the nucleon travels on a curved trajectory. This is not the case in NEUT. But it is an option in ACHILLES and NuWro, and is naturally included in GiBUU.} (in our case, the INC transparency value for calcium at $T_p=100$ MeV, Table~\ref{tab:transparencies}).
Crucially, when the initial nucleon is distributed according to the full nuclear density as is typically done in the two-step approach, one obtains identical transparency for all levels; it would only depends on the nucleon energy, which is not supported by the data.

A similar reasoning applies to the analysis of the JLab data presented in Fig.~\ref{fig:exclusive_12C}. However, due to the larger acceptances and broader missing-energy windows, in this experiment the role of the INC may not be limited to a global reduction in the magnitude of the distributions; it could also slightly modify their shape. 
This is particularly relevant for the deeply bound $1s_{1/2}$ hole state, which contributes at missing energies well above the two-nucleon knockout threshold. Consequently, to draw definitive conclusions, a full simulation incorporating the exact experimental conditions is required.

The relevance of nuclear transparency obtained in the two-step framework to the description of $(e,e'p)$ data is clear.
Indeed, analogously to Eq.~\eqref{eq:T_RDWIA}, a measure of transparency may be defined for each dataset as:
\begin{equation}
\label{eq:T_excl}
    T_\text{model}^\text{exp-PS} = \frac{\int_{\text{exp-PS}} \frac{\de\sigma_\text{ROP}}{\de\Omega_{l}\de E_l\de\Omega_p}}{\int_{\text{exp-PS}}\frac{\de\sigma_\text{model}}{\de\Omega_{l}\de E_l\de\Omega_p}}.
\end{equation}
In this case the integral is over the restricted experimental phase space (exp-PS) of the particular dataset~\footnote{Note that this is not strictly the area under the curves shown in Figs.~\ref{fig:exclusive_40Ca} and \ref{fig:exclusive_12C}, since the cross section is shown as function of $p_m$, but is differential in solid angles. To compute the rates we integrate over $\cos\theta_{p}$ for fixed $q-\omega$ kinematics and over $\cos\theta_l$ for parallel kinematics.}.
In the two-step framework, if the INC reproduces this transparency, the total event rate integrated over  the specific kinematic range under consideration will be the same as that obtained with the ROP.

We have collected the resulting values in Table~\ref{tab:transparencies}. We compare them to the transparencies from Eq.~\eqref{eq:T_RDWIA}, shown in Fig.~\ref{fig:density_transp}, which are computed in a considerably larger phase space.
In Figs.~\ref{fig:exclusive_40Ca}~and~\ref{fig:exclusive_12C}, we have included the EDRMF, r(ROP) and RPWIA predictions multiplied by the spectroscopic factor and the constant transparency factor $T^\text{exp-PS}_\text{model}$.
If the INC reproduces these values of the transparency, one thus obtains the results shown. 
Evidently, to match the shape of the exclusive cross section, instead of only the integrated cross section, no single constant transparency factor will suffice. It is clear that the FSI treatment in the ROP leads to a change of the shape of the cross section, not a simple reduction.

The exclusive transparencies obtained follow the trends observed in Fig.~\ref{fig:density_transp}. There is a clear model dependence, with RPWIA results smaller than those obtained in the EDRMF and r(ROP). This is in particular notable at the lowest proton kinetic energies studied here, i.e. $T_p=100$ MeV.
There is also a clear dependence on the particular shell, the $p$-shell transparency in carbon is consistently larger than that obtained for the $s$-shell. Finally, the transparency decreases with increasing energy and tends to saturate at large $T_p$. 

While these broad trends are the same in the results of Fig.~\ref{fig:density_transp} and those of Figs.~\ref{fig:exclusive_40Ca} and \ref{fig:exclusive_12C}, the exact numerical values differ markedly. 
This is not surprising, the `exclusive' transparency obtained with fixed $\omega-q$ kinematics, when a sufficiently large range in missing momentum is considered, is essentially the same as the ratio of Eq.~\eqref{eq:T_RDWIA} but evaluated at a single fixed momentum transfer $q$.
This ratio can only be $q$-independent if the ROP and real-potential model cross sections at fixed $\omega$ have the same shape.

In summary, the analysis of $(e,e'p)$ data shows that an approach producing transparency that depends only on $T_p$, like the INC, but is independent of other variables (crucially $p_m$ and $q$) will not reproduce the shape of the proton knockout data under exclusive conditions, nor its dependence on specific kinematics. Only the rate integrated over a larger region of phase space can be reproduced in this way.
On the other hand, the shell dependence of the transparency may be readily included in an INC, by sampling the position of the nucleon using the radial density of the state under consideration. 
Furthermore, the transparency required to match $(e,e'p)$ data depends on the specific elementary-vertex model used. Consequently, the INC must be constrained with this model in mind, particularly at low energies.

\begin{table*}[htbp]
\centering
\caption{Transparency ($T$) obtained from different models or definitions. 3th column is the ratio of cross sections,  Eq.~\eqref{eq:T_excl}, over the experimental phase space of the $(e,e'p)$ data and for the EDRMF model; for the calcium data we took the average $T^\text{exp-PS}$ value between the $q-\omega$ const. and parallel kinematics, the exact values are shown in Fig.~\ref{fig:exclusive_40Ca}. 
4th column is as the 3th one but for the RPWIA model. 
5th and 6th columns correspond to the transparency from Eq.~\eqref{eq:T_RDWIA} (Fig.~\ref{fig:density_transp}). 
7th column is the transparency of the NEUT INC defined in Eq.~\eqref{eq:T_rho_int} when the initial proton position is taken to be the total proton density $\rho$, computed with the RMF model (Fig.~\ref{fig:NEUT-transparency}). 
8th column is the transparency from Eq.~\eqref{eq:T_rho_int} for the NEUT INC  when the initial proton position is taken from the shell density $\rho_\kappa$ (Fig.~\ref{fig:density_transp}). 
9th and 10th columns are as the 7th and 8th ones, but with the NuWro INC.}
\label{tab:transparencies}
\vspace{0.2cm}
\begin{tabular}{l|c|c|c|c|c|c|c|c|c}
\hline
\hline
     & $T_p$(MeV) & $T^\text{exp-PS}_{\text{EDRMF}}$ & $T^\text{exp-PS}_\text{RPWIA}$ & $T_\text{EDRMF}$ & $T_\text{RPWIA}$ & $T_\text{NEUT}[\rho]$ & $T_\text{NEUT}[\rho_\kappa]$ & $T_\text{NuWro}[\rho]$ & $T_\text{NuWro}[\rho_\kappa]$ \\ 
\hline\noalign{\hrule height 0.8pt}
${}^{40}\text{Ca}$, $1d_{3/2}$ & $100$ & $0.52$ & $0.45$ & $0.49$ & $0.41$ & $0.40$ & $0.44$ & $0.49$ & $0.53$ \\
\hline
${}^{40}\text{Ca}$, $2s_{1/2}$ & $100$ & $0.64$ & $0.56$ & $0.51$ & $0.43$ & $0.40$ & $0.45$ & $0.49$ & $0.54$ \\ 
\hline\noalign{\hrule height 0.8pt}
                             & $350$ & $0.72$ & $0.71$ & $0.59$ & $0.58$ & $0.57$ &  $0.61$ & $0.62$ & $0.66$\\ 
 ${}^{12}\text{C}$, $1p_{3/2}$ & $700$ & $0.66$ & $0.66$ & $0.51$ & $0.51$ & $0.49$ &  $0.54$ & $0.54$ & $0.59$\\ 
                             & $970$ & $0.62$ & $0.67$ & $0.52$ & $0.52$ & $0.46$ &  $0.52$ & $0.53$ & $0.58$\\ 
\hline
                             & $350$ & $0.55$ & $0.47$ & $0.44$ & $0.42$ & $0.57$ & $0.48$ & $0.62$ & $0.55$\\
${}^{12}\text{C}$,  $1s_{1/2}$ & $700$ & $0.48$ & $0.40$ & $0.34$ & $0.36$ & $0.49$ & $0.39$ & $0.54$ & $0.45$\\
                             & $970$ & $0.52$ & $0.39$ & $0.34$ & $0.37$ & $0.46$ & $0.37$ & $0.53$ & $0.43$\\
\hline
\hline
\end{tabular}
\end{table*}

\section{Discussion and Conclusions}\label{sec:conclusions}

We performed a study of the RDWIA approach and the NEUT and NuWro INC models by comparison to electron scattering data. 
First, we summarize the main results: 
\begin{itemize}
    \item RPWIA fails to reproduce the inclusive $(e,e')$ data at low and intermediate $q$ due to the lack of elastic final-state interactions and Pauli blocking. 
    A real-potential RDWIA model naturally includes the elastic FSI, and better reproduces the QE peak.
    \item A ROP model, with spectroscopic factors, satisfactorily reproduces exclusive $(e,e'p)$ data. 
    \item In the two-step approach of event generators, the single-nucleon knockout cross section (the one of exclusive ${}^{A}\text{X}(e,e'p){}^{A-1}\text{Y}$ scenarios) 
    is given by the elementary-vertex cross section multiplied by a transparency that depends solely on the nucleon energy $T_N$ and is determined by the INC model. 
    \item The nuclear transparency defined as the ratio of ROP to real-potential cross sections and computed over a restricted phase space, the one from the $(e,e'p)$ experimental datasets, does not coincide with the one computed over the full quasi-elastic peak. Consequently, the classical notion of nuclear transparency as a property dependent solely on outgoing kinetic energy breaks down in the RDWIA framework, in contrast to INC models.
    \item A real-potential RDWIA rescaled by a transparency factor, determined so that the cross section integrated over the experimental phase space is the same as those of the ROP result, can qualitatively reproduce $(e,e'p)$ data, but the ROP is superior. The shape of the obtained exclusive cross sections is different.
    \item Nuclear transparency is different for different levels and nuclei, meaning that it depends on (i) the density of the level from which the nucleon originated and (ii) the full nuclear density.
    In the RDWIA approach, information about (i) is in the initial bound-state wave functions, while information about (ii) is in the final scattering-state wave function.
    We have shown that this effect can be reproduced in the classical INC approach.
\end{itemize}

We can now address the questions raised in the introduction:
\begin{itemize}
    \item Which nuclear effects should be included in the initial interaction model and which ones in the hadron propagation?

    Most common nuclear-model approaches in neutrino event generators are based on (infinite nuclear matter) Fermi-gas models, which only account for Fermi motion and, possibly, Pauli blocking; and the PWIA with realistic spectral functions, which can incorporate precise information about the initial state, but ignores key features of the scattering process, like Pauli blocking and elastic FSI. 

    Real-potential RDWIA approaches can account for these nuclear effects and are compatible with INC models, i.e., no effect is double counted if the INC does not contain elastic FSI. 
    We have demonstrated that elastic FSI belong to the elementary-vertex model, they are actually a consequence of using a consistent treatment for the bound and continuum states. 
    Additionally, elastic FSI are necessary to reproduce the inclusive cross section, which is blind to the INC, as well as the exclusive one-nucleon knockout cross section, in which the role of INC is simply to scale down the elementary-vertex prediction by a constant transparency factor.
  
    \item Which experimental data can be used to benchmark and refine both elementary-vertex and INC models?

    For the elementary-vertex model one can use inclusive and exclusive electron scattering. In the latter case, one has to account for the transparency factor to match the magnitude of the data. 

    The two-step approximation can be further constrained with semi-inclusive cross section data, i.e., lepton- and nucleon-detected cross sections that include more than the pure quasifree one-nucleon knockout channel, such as those obtained in experimental facilities with large acceptances~\cite{Dutta03,Khachatryan21} and the recent neutrino-carbon measurement that uses a monochromatic neutrino beam~\cite{Marzec25}. 
    In these cases, one would need to do a full simulation of the experimental conditions, incorporating other possible reaction channels that contribute to the experimental signal. 
    In the particular case of QE scattering, the role of the INC is not only to eliminate events in which the nucleon suffered inelastic interactions, as in an exclusive experiment, but to migrate strength from low to high $E_m$ regions; hence changing the shape, not only the magnitude, of the distributions.

    Traditionally, INC models have been constrained using hadron-nucleus cross sections and nuclear transparency data.
    The goal of event generators is however to reproduce lepton-nucleus scattering data; therefore, $(e,e'p)$ data should be used to test and refine them.  
    This cannot be done by constraining the INC separately from the model used as input. 
    They should be treated consistently, as discussed in the following.

    \item What level of consistency between the elementary-vertex and INC models is required?

    When comparing model predictions with $(e,e'p)$ data below the two-nucleon knockout threshold, within the two-step approach the role of the INC is simply to produce a transparency factor which rescales the cross section obtained in the elementary-vertex model used as input. 
    The nuclear transparency is therefore, by construction, a model-dependent quantity: the transparency produced by the INC should be constrained to the amount needed to match the exclusive electron scattering data; and, as we have shown, the necessary transparency is notably different in RPWIA or real-potential approaches, in particular at low nucleon energies.

    Furthermore, theoretical calculations and comparisons to $(e,e'p)$ show that the transparency depends strongly on the level from which the nucleon is knocked out. 
    We have demonstrated that this effect can be reproduced in the classical INC approach by sampling the initial position of the nucleon according to the density of the shell from which it originated. 
    Current event generators, however, sample the initial position from the total nuclear density instead.
    When using a RDWIA approach in which the contribution from different single-particle states are separated, this feature can be straightforwardly implemented in neutrino event generators.
    This difference in transparency for $s$ and $p$ shell knockout in carbon is shown to be approximately $20\%$. Hence, proper inclusion of this effect is necessary for a comparison to the KDAR data of Ref.~\cite{Marzec25}.
\end{itemize}

We note that while transparency in INC models is independent of the leptonic probe, the transparencies extracted from electron- and neutrino-scattering cross-section ratios are not necessarily identical. These differences have been explored elsewhere~\cite{Martinez06,Butkevich24} and found to be small, however, such discussion is beyond the scope of this work.
    
Finally, the two-step approach discussed here, i.e. `real-potential RDWIA $+$ INC', has intrinsic limitations, consequence of the approximations in which it is based, but its use in neutrino event generators represents an improvement with respect to state-of-the-art approaches currently used in neutrino event generators.

\begin{acknowledgments}
This work was supported by project RYC2022-035203-I and PR47/24-PID2024-160624NB-I00 funded by MCIN/AEI/10.13039/501100011033/FEDER and FSE+, UE;
by ``Ayudas para Atracción de Investigadores con Alto Potencial-modalidad A'' funded by VII PPIT-US; 
by European Union's Horizon research and innovation programme under the Marie Sk\l{}odowska-Curie Actions (MSCA) HORIZON-MSCA-2023-SE-01-01 - MSCA Staff Exchanges 2023, Grant Agreement ID: 101183137 (JENNIFER3 project).
J.~McKean was supported by Grant-in-Aid for JSPS Research Fellows, JSPS KAKENHI Grant No. 25KF0223;
A.N. is supported by the Neutrino Theory Network (NTN) under Award Number DE-AC02-07CH11359; 
T.~Franco-Munoz  was supported by FWO Junior Postdoctoral Fellowship No.~12AG826N.
We acknowledge to RES resources provided by BSC in MareNostrum to project No.~FI-2026-1-0045.
\end{acknowledgments}

\bibliographystyle{apsrev4-1}
\bibliography{bibliography}

\end{document}